\documentclass[11pt,a4paper]{report}
\usepackage{a4}
\usepackage{latexsym}
\usepackage{amsmath}

\begin{document}

\begin{titlepage}

\hspace*{20cm}
\\
\\
\\
\\
\begin{center}

\large{Thesis for the degree of Master of Science in Theoretical Physics}

\hspace{16cm}


\hspace{16cm}

\hspace{16cm}

\huge{Boundary Dynamics of Three-Dimensional Asymptotically Anti-de Sitter Space-Times} \\

\hspace{16cm}

\hspace{16cm}

\large{Sacha van Albada} \\

\hspace{16cm}

Institute for Theoretical Physics \\
Utrecht University \\
December 2003 \\

\flushright

\hspace{16cm}

\hspace{16cm}

\hspace{16cm}

\hspace{16cm}

Under supervision of \\
Prof. B. de Wit \\
Dr. A. E. M. van de Ven \\

\end{center}

\end{titlepage}

\hspace{16cm}

\hspace{16cm}

\begin{center}

\textbf{\huge{Acknowledgments}}

\vspace{0.5cm}

\flushleft

First of all, I would like to thank Bernard de Wit and Anton van de Ven for their patience and for a truly instructive experience. I am indebted to Frank Witte, who helped me settle into the Master's program. 
Mathijs and Siebe showed me how to deal with physics theses. I have been but a poor imitator. I would like to thank my family for their support and never-failing optimism. Finally, I am grateful to Rianne, Jovanka, Linda, and Annemarieke for cheering me up with food many a time!

\end{center}

\newpage

\tableofcontents
\newpage

\chapter{Introduction}

Nature knows of four fundamental forces - electromagnetic, weak, strong, and gravitational. The electrical and magnetic interactions were initially considered as two distinct phenomena, but in the late 19th century, Maxwell discovered a way to view them as two sides of the same coin, and the term electromagnetism was born. The weak and strong interactions describe phenomena at very short distance scales. Gravity, on the other hand, works at such large length scales that we all sense its effect. Since Einstein introduced his theory of general relativity in 1916, it has singled itself out as an accurate description of the gravitational interaction and the curvature of space-time. Gravity is the one fundamental force that has been most difficult to incorporate into a consistent picture of nature. The other three forces have already been combined into a theory called the standard model in the 1970s. The problem with also incorporating gravity is that the standard model is a quantum field theory, but the gravitational field does not allow itself to be quantized in a straightforward way. 

Out of the four fundamental interactions, we will mainly be concerned with gravity. However, we will not be working in the $3+1$ (three spatial, one time) dimensions of our physical world, but rather in $2+1$ dimensions, where gravity is simplified because it does not possess any local propagating degrees of freedom. What makes the theory interesting after all are global properties \cite{Deser}, and a phenomenon we call `boundary dynamics.' Based on a work by J. D. Brown and M. Henneaux (1986) \cite{Brown-Henneaux}, we will look at the group of asymptotic symmetries of a class of asymptotically anti-de Sitter space-times. Anti-de Sitter space-time is the maximally symmetric solution of the Einstein equations with constant negative curvature and no matter sources. `Asymptotically anti-de Sitter' means that we require the space-time to behave like anti-de Sitter space-time near spatial infinity. The presence of the boundary partially breaks diffeomorphism invariance, causing only certain transformations (asymptotic symmetries) to remain as symmetries. 

\newpage

Brown and Henneaux noticed that the asymptotic symmetry algebra is special in two respects. For one, the AdS$_3$ isometry group is extended to an infinite-dimensional group at the boundary, namely the group of conformal transformations of the plane. Since the conformal group has only a finite number of generators in higher dimensions, this is a phenomenon specific to the $2+1$-dimensional theory. Besides the fact that the asymptotic symmetries form an infinite-dimensional group, Brown and Henneaux noticed that their Poisson bracket algebra is \emph{centrally extended}. They also calculated the associated central charge. The fact that the central extension arises already at the classical level makes it even more special, because central extensions are best known as a quantum effect. In this thesis it will become clear that central extensions may very well arise classically.

The asymptotic symmetry group is generated in the Hamiltonian formalism by two copies of the Virasoro algebra, suggesting a dual interpretation in terms of a conformal field theory on the $(1+1)$-dimensional boundary at spatial infinity. It was shown in \cite{Coussaert} that this conformal field theory is Liouville theory. A supersymmetric generalization of this was presented in \cite{AdS supergravity,Bautier} for the case $\mathcal{N} = 1$ and in \cite{Maoz} for extended supergravity. These results are reminiscent of the AdS/CFT correspondence which was conjectured by Maldacena in 1998 \cite{Maldacena}, and concerns an analogy between $(d-1)$-dimensional supersymmetric conformal field theory and black branes of which the near-horizon geometry is AdS$_d$ times a compact manifold.

The conformal field theory interpretation has led to another interesting result. There is a well-known formula due to Cardy \cite{Cardy} relating the central charge of two-dimensional conformal field theory to the density of high-energy states in the theory. The logarithm of the density of states gives an expression for the entropy. Strominger \cite{Strominger} has shown that, using the central charge of asymptotically AdS$_3$ space-times, Cardy's formula yields exactly the Bekenstein-Hawking \cite{Bekenstein, Hawking} entropy of a black hole solution discovered by Ba\~nados, Teitelboim and Zanelli (BTZ) in 1992 \cite{BTZ}. The BTZ black hole is a three-dimensional solution of Einstein's equations with negative cosmological constant and a point source at the origin. Far away from the source, the solution behaves like anti-de Sitter space, and the BTZ black hole metric is asymptotically anti-de Sitter in the sense of \cite{Brown-Henneaux}. Generalizations of Strominger's approach to different solutions have been suggested for instance in \cite{Killing horizons}.

We have performed our analysis in the metric formulation, following Brown and Henneaux. An alternative would have been the Chern-Simons formulation of $(2+1)$-dimensional gravity \cite{Witten}, which is a gauge theory in terms of a one-form field $A_i$ taking values in the space-time isometry group. The specific case of (asymptotically) AdS$_3$ was studied in the Chern-Simons formulation in \cite{Ortiz} and \cite{Banados}. The derivations of the conformal field theories on the boundary in \cite{Coussaert} and \cite{Maoz} also make use of this formulation, of which the supersymmetric generalization was given in \cite{Achucarro}.

\newpage

This thesis is organized as follows. In Chapter 2, some preliminaries are given on isometries and conformal symmetries, and we become familiar with the Virasoro algebra. Two examples of classical central charges are discussed. Chapter 3 contains an introduction to the Hamiltonian formulation of gauge theories in the context of Maxwell theory. The knowledge gained in Chapter 3 is applied to general relativity in Chapter 4. The Hamiltonian is shown to acquire a surface term due to the presence of the boundary. The geometrical properties of anti-de Sitter space and the BTZ black hole are the subject of Chapter 5. The main part of the discussion follows in Chapter 6, which contains the calculation of the central charge in the asymptotic symmetry algebra of asymptotically AdS$_3$ space-times. After some preliminaries on the AdS/CFT correspondence and Chern-Simons theory, the derivation of the boundary conformal field theory \cite{Coussaert} is summarized in Chapter 7. Finally, Strominger's entropy calculation is presented in Chapter 8, along with some comments on this approach.

Throughout, our metrics will have signature $(-+\ldots+)$, where the first entry denotes the time component.

\chapter{Isometries and Conformal Symmetry}
In this chapter we introduce the notions of isometries and conformal symmetry. We motivate their introduction by considering the symmetry properties of the scalar field theory defined by the action
\begin{equation}\label{scalar field theory}
S = \int \sqrt{-g(x)}g^{\mu\nu}(x)\partial_{\mu}\phi(x)\partial_{\nu}\phi(x)
 ~d^{d}x.
\end{equation}
Section 2.3 deals with the central extension of the conformal algebra in two dimensions, also known as the Virasoro algebra. In section 2.4, we discuss a few examples of theories that are centrally extended at the classical level.

\section{Coordinate Transformation}
First consider the theory (\ref{scalar field theory}) in arbitrary dimension $d$. We define
the Lagrangian density $\mathcal{L}(g, \phi)$ to be the entire integrand,
including the factor $\sqrt{-g(x)}$. The equation of motion is
\begin{equation}
D^{\mu}\partial_{\mu}\phi = 0.
\end{equation}
The metric is not treated as a dynamical field here, because our
action does not contain an Einstein-Hilbert term. Since on scalar
fields the covariant derivative is simply
the partial derivative, we can still expect there to be some
equivalence between initial and transformed Lagrangians under
general diffeomorphisms. We will look exactly what happens under
the infinitesimal transformation $x^{\mu} \to x^{\mu} -
\xi^{\mu}(x)$. Performing an active transformation, such that we
describe the new fields in terms of the old coordinates $x^{\mu}$,
we have
\begin{eqnarray}\label{variations}
\delta\sqrt{-g} &=& \frac{1}{2}\delta g^{\mu}_{~\mu}\sqrt{-g} =
(\frac{1}{2}g^{\mu\nu}\xi^{\rho}\partial_{\rho}g_{\mu\nu} + \partial_{\rho}\xi^{\rho})\sqrt{-g}, \nonumber \\
\delta\partial_{\mu}\phi &=&
\partial_{\mu}(\xi^{\nu}\partial_{\nu} \phi) =
\partial_{\mu}\xi^{\nu}\partial_{\nu}\phi +
\xi^{\nu}\partial_{\nu}\partial_{\mu}\phi, \nonumber \\
\delta g^{\mu\nu} &=& \xi^{\rho}\partial_{\rho}g^{\mu\nu} -
g^{\rho\nu} \partial_{\rho}\xi^{\mu} - g^{\rho\mu}\partial_{\rho}\xi^{\nu}.
\end{eqnarray}
Notice that by $\delta g^{\mu}_{~\mu}$ we mean the trace of $\delta g_{\mu\nu}$,
not the variation of the trace of $g_{\mu\nu}$, which would be $\xi^{\rho}\partial_{\rho}g^{\mu}_{~\mu}$.
It can be verified that under these transformations,
\begin{equation}\label{total derivative}
\delta \mathcal{L} = \partial_{\nu}(\xi^{\nu} \mathcal{L}),
\end{equation}
so that we obtain the equivalence statement that under
infinitesimal general coordinate transformations, $\mathcal{L}(g, \phi)$
goes into a new Lagrangian $\mathcal{L}(g', \phi')$ which has the same
equations of motion but in terms of different fields. This is not
a true invariance, as the background metric has changed in the
process. Because the metric is not a dynamical field in our example,
we have to choose a specific metric beforehand instead of determining
it from the equations of motion. Therefore, the next thing to do is to look at invariances of the metric (isometries), generated by so-called Killing
vectors. This will lead to a transformed Lagrangian $\mathcal{L}(g, \phi')$
describing the same theory as $\mathcal{L}(g, \phi)$, only the scalar field
having changed.

Under the infinitesimal transformation $x^{\mu} \to x^{\mu} -
\xi^{\mu}(x)$, the metric changes by the Lie derivative (as we
already used in contravariant form in (\ref{variations})), which
can be rewritten as a sum of covariant derivatives of the
deformation vector:
\begin{eqnarray}\label{Lie derivative}
g'_{\mu\nu}(x) - g_{\mu\nu}(x) &=& \mathcal{L}_{\xi}g_{\mu\nu} \nonumber \\
&=& \xi^{\rho}\partial_{\rho}g_{\mu\nu}+ g_{\nu\rho}\partial_{\mu}\xi^{\rho} +
g_{\mu\rho}\partial_{\nu}\xi^{\rho}
 \nonumber \\
&=& \partial_{\mu}(g_{\nu\rho}\xi^{\rho}) +
\partial_{\nu}(g_{\mu\rho}\xi^{\rho}) -
2\Gamma^{\beta}_{~\mu\nu}\xi_{\beta} \nonumber \\
&=& \partial_{\mu}\xi_{\nu} - \Gamma^{\beta}_{~\mu\nu}\xi_{\beta}
+
\partial_{\nu}\xi_{\mu} - \Gamma^{\beta}_{~\nu\mu}\xi_{\beta}
\nonumber \\
&=& D_{\mu}\xi_{\nu} + D_{\nu}\xi_{\mu}
\end{eqnarray}
where we assumed the connection to be the Christoffel connection
\begin{equation}
\Gamma^{\beta}_{~\mu\nu} =
\frac{1}{2}g^{\beta\alpha}(\partial_{\mu}g_{\alpha\nu} +
\partial_{\nu}g_{\alpha\mu} - \partial_{\alpha}g_{\mu\nu}),
\end{equation}
which makes sure that $D_{\rho}g_{\mu\nu} = 0$. From (\ref{Lie
derivative}) it follows that Killing vectors $\xi^{\mu}$ obey
\begin{equation}
D_{\mu}\xi_{\nu} + D_{\nu}\xi_{\mu} = 0,
\end{equation}
an equation with at most $\frac{1}{2}d(d+1)$ linearly independent
solutions. For instance, flat Minkowski space is a maximally
symmetric space, its Killing vectors corresponding to the
$\frac{1}{2}d(d+1)$ generators of the Poincar\'e group
($\frac{1}{2}d(d-1)$ Lorentz transformations and $d$
translations).

The variation of the Lagrangian is proportional to the field equations under any transformation of $\phi$. Since the action is invariant under diffeomorphisms, the part of the variation proportional to $\delta g_{\mu\nu}$ is therefore independently proportional to the field equations for any diffeomorphism $\xi^{\mu}$. If we introduce the energy-momentum, or stress, tensor
\begin{equation}
T_{\mu\nu} = - \frac{2}{\sqrt{-g}}\frac{\delta \mathcal{L}}{\delta
g^{\mu\nu}},
\end{equation}
we can denote this part of the variation by
\begin{equation}\label{prop to g}
\delta S_g = - \int \frac{1}{2} \sqrt{-g} ~T^{\mu\nu} \delta g_{\mu\nu} d^d x.
\end{equation}
Because $\delta g_{\mu\nu} = D_{\mu}\xi_{\nu} + D_{\nu}\xi_{\mu}$,
and $T^{\mu\nu}$ is symmetric in its indices, we can integrate by
parts to get
\begin{equation}
\delta S_g = \int \sqrt{-g} ~\xi_{\nu}D_{\mu}T^{\mu\nu} d^d x
\end{equation}
Since $\xi_{\nu}$ is arbitrary, this shows that $T^{\mu\nu}$ is
covariantly conserved, $D_{\mu}T^{\mu\nu} = 0$, in accordance with
Noether's theorem that every continuous symmetry of the action brings
along a conserved current.

In our present example, using $\delta\sqrt{-g} = -\frac{1}{2}\sqrt{-g}~g_{\mu\nu}\delta g^{\mu\nu}$,
it can be verified that (\ref{prop to g}) gives
\begin{equation}\label{current}
T_{\mu\nu} = g^{\rho\sigma}\partial_{\rho}\phi
\partial_{\sigma}\phi g_{\mu\nu}- 2\partial_{\mu}\phi
\partial_{\nu}\phi.
\end{equation}
It can be checked that $T_{\mu\nu}$ is covariantly conserved if the scalar field obeys its equation of motion.

The conserved (divergenceless) current $T_{\mu\nu}$ implies the
existence of $d$ conserved charges
\begin{equation}
Q_{\mu} = \int T^{0}_{~\mu} ~d^{d-1}x
\end{equation}
where the index $0$ stands for time and we integrate over the spatial directions. Indeed, we have
\begin{eqnarray}
\partial_{0} Q_{\mu} &=& \int \partial_0 T^0_{~\mu} ~d^{d-1}x \nonumber \\
&=& - \int \partial_{i} T^{i}_{~\mu} ~d^{d-1}x,
\end{eqnarray}
which vanishes by virtue of Gauss's theorem assuming $T^{i}_{\mu}$ vanishes at the boundaries (a Latin index denotes spatial components).

\section{Conformal Symmetry}
In general, a theory can have more symmetries than only those
generated by the Killing vectors. We could for instance try to
rescale the metric by a local factor (a \emph{Weyl} rescaling),
$g_{\mu\nu}(x) \to f(x)g_{\mu\nu}(x)$. One example for which this
is a symmetry is four-dimensional Maxwell theory, for which the
Lagrangian in terms of the antisymmetric field strength tensor
$F_{\mu\nu} = \partial_{\mu}A_{\nu} -
\partial_{\nu}A_{\mu}$ reads
\begin{equation}\label{Maxwell theory}
\mathcal{L} = -\frac{1}{4} \sqrt{-g} ~g^{\alpha\mu}g^{\beta\nu}F_{\alpha\mu}F_{\beta\nu}.
\end{equation}
With the Lagrangian written in this form, it is easy to see that
in four dimensions the theory is invariant under Weyl rescalings,
since $g^{\mu\nu}(x) \to \frac{1}{f(x)}g^{\mu\nu}(x)$, and in four
dimensions $\sqrt{-g}$ transforms with $f^2(x)$.

It is sometimes also possible to achieve a rescaling of the metric
through diffeomorphisms $x^{\mu} \to x^{\mu} - \xi^{\mu}(x)$, a
\emph{conformal} transformation. Four-dimensional source-free
Maxwell theory in flat space-time is conformally invariant. As
soon as a source term $A_{\mu}J^{\mu}$ is added to the Lagrangian,
conformal invariance is lost. Besides four-dimensional Maxwell
theory, there are many more examples of conformally invariant
theories. Nonlinear sigma models in flat space-time and with
conical target space can be made scale invariant in any dimension
by adding appropriate improvement terms \cite{de Wit}. Mass terms
generally break scale invariance.

A field $\Psi$ is called conformally invariant with conformal weight or conformal dimension $w$ if $\Psi$ is a solution of the field equations with metric $g_{\mu\nu}$, and $\Omega^w\Psi$ is a solution with metric $\Omega^2 g_{\mu\nu}$. Fields obtained by functional differentiation of a conformally invariant action with respect to the metric are conformally invariant. In particular, the stress tensor of a conformally invariant theory is conformally invariant. For tensor fields, the conformal weight depends on the index positions. 

The vectors achieving a rescaling of the metric obey the equation
\begin{equation}\label{conformal Killing equation}
D_{\mu}\xi_{\nu} + D_{\nu}\xi_{\mu} =
\frac{2}{d}g_{\mu\nu}D_{\rho}\xi^{\rho}.
\end{equation}
Vectors obeying (\ref{conformal Killing equation}) are called
\emph{conformal} Killing vectors. They form a group, just like the
Killing vectors. From the right-hand side it can be seen that they
indeed rescale the metric by a local factor, since the left-hand
side is the variation of $g_{\mu\nu}$ under subtraction of
$\xi^{\mu}$ from the coordinates. Whereas the Killing equations
require $D_{\mu}\xi_{\nu} + D_{\nu}\xi_{\mu}$ to be traceless, the
conformal Killing equations lead to no such restriction. Therefore
there are generally more conformal Killing vectors than there are
Killing vectors. Indeed, in more than two dimensions, the
conformal Killing vectors of flat $d$-dimensional Minkowksi
spacetime generate the so-called conformal group $SO(d,2)$,
consisting of $d$ translations, $\frac{1}{2}d(d-1)$ Lorentz
transformations, one dilatation and $d$ so-called special
conformal transformations. The total conformal group therefore has
$\frac{1}{2}(d+1)(d+2)$ generators, whereas the Killing vectors
only corresponded to the $\frac{1}{2}d(d+1)$ generators of the
Poincar\'e group. The transformations of the conformal group in
more than two dimensions and their infinitesimal forms are
summarized in Table 2.1.
\\
\begin{table}[!h]
\label{conformal group}
\begin{tabular}{l|c|l}
Operation & Finite Form & Generator \\
\hline
& & \\
translations & $x'^{\mu} = x^{\mu} + \lambda^{\mu}$ & $P_{\mu} = \partial_{\mu}$ \\
& & \\
Lorentz transformations & $x'^{\mu} = M^{\nu}_{\mu} x^{\nu}$ & $L_{\mu\nu} = x_{\mu}\partial_{\nu} - x_{\nu}\partial_{\mu}$ \\
& & \\
dilatations & $x'^{\mu} = A x^{\mu}$ & $D = -x^{\mu}\partial_{\mu}$ \\
& & \\
special conformal transformations & $x'^{\mu} = \frac{x^{\mu} - x^2 b^{\mu}}{1 - 2x^{.}b - x^2 b^2}$ & $K_{\mu} = x^2\partial_{\mu} - 2x_{\mu}x^{\nu}\partial_{\nu}$ \\
& & \\
\hline
\end{tabular}
\caption{The conformal group for $d>2$.}
\end{table}
\\
We can define the new generators
\begin{eqnarray}\label{SO(2,2) generators}
G_{-1,\mu} &=& \frac{1}{2}(P_{\mu} - K_{\mu}), \nonumber \\
G_{0,\mu} &=& \frac{1}{2}(P_{\mu} + K_{\mu}), \nonumber \\
G_{-1,0} &=& D, \nonumber \\
G_{\mu\nu} &=& L_{\mu\nu},
\end{eqnarray}
where we introduced two extra indices, so that Latin indices take
on the values $-1, 0, \ldots ,d$, while Greek indices run over $1,
2, \ldots, d$. In terms of the generators $G_{ab}$, $SO(d,2)$ is
described by the algebra
\begin{equation}
[G_{ab}, G_{cd}] = \eta_{ac}G_{bd} + \eta_{db}G_{ac} - \eta_{ad}G_{bc} - \eta_{bc}G_{ad}
\end{equation}
with $\eta_{\mu\nu} = diag(-1, -1, 1, \ldots, 1)$.

For $d \ge 3$, the conformal group in $d$ dimensions corresponds
to the anti-de Sitter group in $d+1$ dimensions. The anti-de
Sitter group is the group of isometries of anti-de Sitter space.
We will come back to this in Chapter 5.

We already know that the number of linearly independent conformal
Killing vectors does not exceed $\frac{1}{2}(d+1)(d+2)$ for
dimensions greater than two. However, in two dimensions, the
conformal group is extended to an infinite-dimensional group. To
see this, we turn to holomorphic coordinates $z, \bar z$, for
which the metric has only the components $g_{z\bar z} = g_{\bar z
z} = 1$\footnote{The `holomorphic' coordinates will be complex in
the Euclidean case, but real in ($1+1$)-dimensional Minkowski
space. In the latter case, $z = \sigma + \tau$, $\bar z = \sigma -
\tau$, with $\sigma$ the spatial and $\tau$ the time coordinate,
and $z$ and $\bar z$ are more commonly referred to as lightcone
coordinates.}. This can be done because any two-dimensional metric
is conformally flat. With the given form of the metric, it can be
checked that each of the vectors
\begin{equation}
l_n = -z^{n+1}\partial_z,\ \ \ \bar l_n  = -\bar z^{n+1}\partial_{\bar z},
\end{equation}
$n \in Z$, satisfies the conformal Killing equations (\ref{conformal Killing equation}).
Hence there is an infinite number of conformal Killing vectors in $d = 2$. The vectors
$l_n$ obey the so-called loop, or Witt, algebra
\begin{equation}\label{loop}
[l_m, l_n] = (m-n)l_{m+n},
\end{equation}
and similarly for $\bar l_n$, whereas the $l_n$ and $\bar l_m$
commute among each other. The term loop algebra derives from the fact that if one sets $z = e^{i\phi}$, (\ref{loop}) describes the algebra of vector fields on a circle in the complex plane. The six vectors $l_{-1}, l_0, l_1, \bar
l_{-1}, \bar l_0, \bar l_1$ generate a subgroup $SL(2,R)\times
SL(2,R)/Z_2$ isomorphic to $SO(2,2)$. This group is often written
as the product of a left-hand and a right-hand group
$SL(2,R)_L\times SL(2,R)_R$.

Now consider the theory (\ref{scalar field theory}) in $d = 2$.
Because in two dimensions $g$, the determinant of $g_{\mu\nu}$,
transforms with the inverse square of $g^{\mu\nu}$, we see that
the Lagrangian is invariant under local rescalings of the
metric. So two is a special number of dimensions in this respect.

Associated with the conformal invariance is again a conserved Noether current $J_{\mu}$, which now looks like
\begin{equation}
J_{\mu} = T_{\mu\nu}\xi^{\nu},
\end{equation}
where $T_{\mu\nu}$ is the energy-momentum tensor specified before, and $\xi^{\nu}$ a conformal Killing vector. $J_{\mu}$ is covariantly conserved by virtue of
\begin{eqnarray}
D_{\mu}J^{\mu} &=& \xi^{\nu}D_{\mu}T^{\mu}_{~\nu} + T^{\mu}_{~\nu}D_{\mu}\xi^{\nu} \nonumber \\
&=& \frac{1}{2}T^{\mu\nu}(D_{\mu}\xi_{\nu} + D_{\nu}\xi_{\mu}) \nonumber \\
&=& \frac{1}{d}T^{\mu}_{~\mu}D_{\rho}\xi^{\rho} \nonumber \\
&=& (1 - \frac{2}{d})\partial^{\mu}\phi\partial_{\mu}\phi D_{\rho}\xi^{\rho}.
\end{eqnarray}
The last expression obviously vanishes in the case $d=2$. The
third line immediately shows that conformal invariance implies the
vanishing of the trace of the energy-momentum tensor,
$T^{\mu}_{~\mu} = 0$.

According to a well-established theorem, the stress tensor
classically generates transformations of the canonical variables
through the Poisson bracket. Let us consider this in two
dimensions. In holomorphic coordinates $T_{z\bar z}(z, \bar z)$
vanishes, and $\bar T_{zz}(\bar z) = T_{\bar z \bar z}(\bar z)$,
so that it has become common practice to denote $T_{zz}(z)$ simply
by $T(z)$. The latter can then be expanded in terms of an infinite
number of operators $L_{n}$ according to
\begin{equation}\label{stress}
T(z) = \sum_{n \in Z} L_{n}z^{-n-2},
\end{equation}
and similarly for $\bar T(\bar z)$ in terms of $\bar L_{n}$. The
components $L_{n}$ of the energy-momentum tensor $T_{\mu\nu}$ will
generate symmetries of the two-dimensional Lagrangian, just like
$T_{\mu\nu}$ itself, and hence will form an algebra. On the classical configuration space, the $L_{n}$ defined by
(\ref{stress}) turn out to obey the same algebra as the vectors
$l_n$, namely
\begin{equation}
\{L_m, L_n\} = (m-n)L_{m+n}
\end{equation}
So the symmetry group of the scalar field theory
(\ref{scalar field theory}) in two dimensions is generated by two
copies of the loop algebra discussed above. In some theories a modified version of
this algebra arises, and this is discussed in the next section.

\section{Central Extension}
We are still looking at the infinite-dimensional group of
conformal transformations that arises in $d = 2$. We mentioned
that the algebra (\ref{loop}) may be modified in some cases. The
modified algebra is called a \emph{central extension} of the loop
algebra. 

When quantizing the two-dimensional theory, the components $L_{n}, \bar L_n$ of the energy-momentum tensor become operators on the quantum mechanical
Hilbert space. They can then be expanded in terms of products of
raising and lowering operators. After commuting two different
$L_m$, $L_n$, these raising and lowering operators will be mixed
up, and since they do not commute among each other, the result
will not immediately be recognizable as another $L_{n'}$ or $\bar
L_{n'}$. In order to close the algebra after all, we need to give
a prescription for the order in which to put the raising and
lowering operators in the expansion. This is called normal
ordering. Conventionally, all raising operators are written to the
left of the lowering operators. This leads to an additional term
in the algebra which commutes with all the generators and is
called `central' for that reason. However, normal ordering is not the only way in which to obtain a central charge. In Chapter 6 we will come across a central extension of the Virasoro algebra associated with asymptotic isometries of a class of
asymptotically $AdS_3$ spacetimes which is not
quantum in nature. A classical central charge was first described
by Gervais and Neveu in the context of Liouville
theory~\cite{Gervais}. More on the topic of classical central
charges is explored in section 2.4.

The central extension of the loop
algebra, the \emph{Virasoro} algebra, is best known from string
theory. For string theory, two is a special number of dimensions,
since it is the dimension of the worldsheet swept out by the
strings, and string theory has a $d=2$ conformal invariance.

Let us have a look at the relation between the quantum mechanical,
centrally extended loop algebra and its classical limit. In order
to pass to quantum mechanics, we first replace the Poisson bracket
by $\frac{1}{i\hbar}$ times the commutator:
\begin{equation}
[L_m, L_n] = i\hbar(m-n)L_{m+n}.
\end{equation}
Defining $L'_m \equiv \frac{L_m}{i \hbar}$, we have
\begin{equation}
[L'_m, L'_n] = (m-n)L'_{m+n}.
\end{equation}
Performing the normal ordering leads to (we denote normal ordered operators by a script letter)
\begin{equation}
[\mathcal{L}'_m, \mathcal{L}'_n] = (m-n)\mathcal{L}'_{m+n} + \mathrm{central ~term}.
\end{equation}
Such a centrally extended algebra has so-called \emph{projective respresentations}.
We can formally take the classical limit by going back to $\mathcal{L}_m \equiv i \hbar \mathcal{L}'_m$, obtaining
\begin{equation}\label{quantum central extension}
[\mathcal{L}_m, \mathcal{L}_n] = i\hbar\Big( (m-n)\mathcal{L}_{m+n} + i \hbar* \mathrm{central ~term}\Big),
\end{equation}
subsequently replacing the commutator by $i\hbar$ times the
Poisson bracket, and taking the limit $\hbar \to 0$. The central
term then drops out.

By making use of the Jacobi identity and through addition of constants
to the generators it can be shown that the central
extension can always be written in the form (henceforth dropping
primes)
\begin{eqnarray}\label{central extension}
[\mathcal{L}_{m}, \mathcal{L}_{n}] &=& (m - n)\mathcal{L}_{m+n} + \frac{c}{12}(m^3 - m)\delta_{m+n,0},  \nonumber \\
\left[ \bar{\mathcal{L}}_{m}, \bar{\mathcal{L}}_{n} \right] &=& (m-n)\bar{\mathcal{L}}_{m+n} + \frac{\bar{c}}{12}(m^3 - m)\delta_{m+n,0}, \nonumber \\
\left[\mathcal{L}_{m}, \bar{\mathcal{L}}_{n} \right] &=& 0,
\end{eqnarray}
where $c$ is an undetermined central charge (the factor $\frac{1}{12}$ is
according to convention). The commutation relations (\ref{central
extension}) are those of two copies of the Virasoro algebra with
central charge $c$. The notation is somewhat misleading, since,
whereas the $\mathcal{L}_n$ can be interpreted as generators, the central
term is just a constant of which it is not clear what it generates.
The generators $\mathcal{L}_{-1}, \mathcal{L}_0, \mathcal{L}_1,
\bar{\mathcal{L}}_{-1}, \bar{\mathcal{L}}_0, \bar{\mathcal{L}}_1$
form a subalgebra isomorphic to $so(2,2)$ without central
extension, which is obvious with the $m$-dependence in
(\ref{central extension}).

The central charge also shows up in the transformation law of the
two-dimensional stress tensor. Using the components of the stress
tensor as generators of transformations through the Lie
bracket, we can derive the infinitesimal variation of, in this
case, the stress tensor itself. Of course this works not only for
$T(z)$, but for any function on the Hilbert space. Afterwards, we
also give the finite form of the variation.

Just like the $l_n$, $\mathcal{L}_{n}$ generates
the transformation $z \to z - \epsilon z^{n+1}$, and (leaving out the infinitesimal parameter $\epsilon$) we have
\begin{equation}
\delta_{z^{n+1}}T(z) = [\mathcal{L}_{n}, T(z)].
\end{equation}
Again, $T(z)$ can be expanded in terms of the Virasoro operators as
\begin{equation}
T(z) = \sum_{m \in Z} \mathcal{L}_{m}z^{-m-2},
\end{equation}
and we get
\begin{eqnarray}
\delta_{z^{n+1}}T(z) &=& [\mathcal{L}_n,\sum_{m \in Z} \mathcal{L}_{m}z^{-m-2}] \nonumber \\
&=& \sum_{m \in Z}z^{-m-2}((n-m)\mathcal{L}_{m+n} + \frac{c}{12}n(n^2 -
1)\delta_{n+m,0}) \nonumber \\
&=& (z^{n+1}\partial_{z} + 2\partial_{z}z^{n+1})\sum_{m \in
Z}\mathcal{L}_{m}z^{-m-2} + \frac{c}{12}\partial_{z}^{3}z^{n+1}.
\end{eqnarray}
The same can be done for general $n$, so that we arrive at (from
now on denoting $\partial_{z}$ simply by $\partial$)
\begin{equation}\label{nontensorial}
\delta_{\xi}T(z) = (\xi\partial + 2\partial\xi)T(z) +
\frac{c}{12}\partial^{3}\xi.
\end{equation}
The first part is tensorial (i.e. it is according to the tensor transformation law $T'^{\mu\nu}(x') = \frac{\partial x'^{\mu}}{\partial x^{\rho}}\frac{\partial x'^{\nu}}{\partial x^{\sigma}}T^{\rho\sigma}(x)$), in contrast to the second part, which
features the same central charge as appears in the central extension of
the Virasoro algebra.

We will now look at the finite form of the transformation
(\ref{nontensorial}). If $f(z) = z + \xi(z) + O\left(\xi^2\right)$, the transformation law
(\ref{nontensorial}) is the infinitesimal version of
\begin{equation}\label{nonlinear}
T(z) = (\partial f)^{2} T(f(z)) + \frac{c}{12}\{f,z\}
\end{equation}
where the Schwarzian derivative $\{f,z\}$ is defined by
\begin{equation}
\{f,z\} = \frac{\partial f \partial^{3}f -
\frac{3}{2}(\partial^{2}f)^2}{(\partial f)^2}.
\end{equation}
This can be seen by calculating (\ref{nonlinear}) up to first
order in $\xi$. 

\section{Classical Central Charges}

In the above discussion we have treated the central term as a
quantum effect, which it is in most cases. However, we have
mentioned that central charges may perfectly well arise already at the
classical level. This section introduces the concept of classical central
charges by means of two examples. One is a scalar field theory in $(1+1)$ dimensions with constant external electric field, in which the center of the symmetry algebra actually generates a transformation. The other is Liouville theory, where the central term only shows up at the level of the Poisson brackets, and does not act as a generator. We will see that there is a difference in the mechanisms by which these two central terms arise. The possibility for a central charge that does not generate a transformation derives from an ambiguity in the canonical generators. The central charge in the first example, on the other hand, is shown to arise due to a partially broken symmetry that is restored by additional symmetries. 

Our first example of a classical central extension was described by E. Karat in
\cite{Karat}. It concerns a charged scalar field theory in flat
$(1+1)$-dimensional space with a constant external electric field. The Lagrangian is
\begin{equation}\label{charged scalar field theory}
\mathcal{L} = -(D_{\mu}\phi)^* (D^{\mu}\phi) - m^2\phi^*\phi
\end{equation}
with
\begin{eqnarray}
D_{\mu}\phi = \partial_{\mu}\phi + ieA_{\mu}\phi \nonumber \\
A_{\mu} = -\frac{1}{2}\epsilon_{\mu\nu}x^{\nu}F,
\end{eqnarray}
and $\epsilon^{10} = -\epsilon_{10} = 1$. The electric field $E$ then becomes
\begin{equation}
E = \partial_0 A_1 - \partial_1 A_0 = F,
\end{equation} 
and it does not point in any direction. Due to the lack of a dynamical term for $A_{\mu}$, the vector potential does not transform, and the action is not manifestly invariant under translations. If we pretended the $A_{\mu}$-field to transform after all, the total symmetry group would be a product of the Poincar\'e group, consisting of Lorentz transformations and translations
\begin{eqnarray}
\phi &\to& \phi + t^{\mu}\partial_{\mu}\phi \nonumber \\
A_{\mu} &\to& A_{\mu} -\frac{1}{2}\epsilon_{\mu\nu}F t^{\nu} \nonumber \\
\delta_{T}\mathcal{L} &=& \delta_{T_{\phi}}\mathcal{L} + \delta_{T_{A}}\mathcal{L} 
\nonumber \\
&=& t^{\mu}\partial_{\mu}\mathcal{L}, 
\end{eqnarray}
and gauge transformations
\begin{eqnarray}\label{transform gauge}
\phi &\to& e^{ie\Lambda}\phi \nonumber \\
A_{\mu} &\to& A_{\mu} - \partial_{\mu}\Lambda \nonumber \\
\delta_{g}\mathcal{L} &=& \delta_{g_{\phi}}\mathcal{L} + \delta_{g_{A}}\mathcal{L} \nonumber \\
&=& 0.
\end{eqnarray}
However, the change in $A_{\mu}$ under translations is just a gauge tranformation
\begin{eqnarray}
\delta_T A_{\mu} &=& -\frac{1}{2}\epsilon_{\mu\nu}F t^{\nu} \nonumber \\
&=& \partial_{\mu}(-\frac{1}{2}\epsilon_{\nu\rho}x^{\rho}t^{\nu}),
\end{eqnarray}
as we could have guessed, because the electric field is itself translation invariant. From (\ref{transform gauge}), we then see that we can `translate' the transformation of $A_{\mu}$ under translations into an infinitesimal gauge transformation on $\phi$, obtaining a so-called \emph{covariant translation}:
\begin{equation}\label{strange translations}
\delta \phi = t^{\mu}(\partial_{\mu} + \frac{1}{2}ie\epsilon_{\mu\nu}x^{\nu}F)\phi.
\end{equation}
Defining the transformation of $\phi$ under translations to be (\ref{strange translations}), we recover translation invariance of the action when $A_{\mu}$ does not transform,
\begin{equation}
\delta_T \mathcal{L} = t^{\mu} \partial_{\mu}\left[(-D^{\nu}\phi)^* D_{\nu}\phi -m^2\phi^* \phi \right].
\end{equation}
However, the covariant translations do not commute like regular translations do,
\begin{eqnarray}
\left[ \delta_L, \delta^{\mu}_T\right] \phi = \epsilon^{\mu\nu} (\delta_{T})_{\nu} \phi \nonumber \\
\left[ \delta^{\mu}_{T}, \delta^{\nu}_{T} \right] \phi = ie\epsilon^{\mu\nu}F \phi.
\end{eqnarray}
For simplicity, we have left out the translation parameters. The operator that multiplies with $ieF$ is a central element, since it commutes with the other symmetry
operations. The same algebra is obeyed by the corresponding
Noether charges, but the central term $ie\epsilon^{\mu\nu}F$
already shows up before introducing Poisson brackets.
The above example gives a good idea of how central charges may
show up already at the classical level. 

In our next example, the center does not act as a generator. The possibility for such a central term in a Poisson bracket algebra can be explained as follows, with an argument from
\cite{Toppan}. Suppose the symmetries of the action close
according to
\begin{equation}\label{closure}
[\delta_a,\delta_b]\phi = f_{ab}^{~~c} \delta_c \phi,
\end{equation}
and we have a set of Noether charges generating these symmetries,
\begin{equation}
\delta_a \phi = \{Q_a,\phi \},
\end{equation}
where $\{ \;  , \; \}$ denotes the Poisson bracket,
\begin{equation}\label{Poisson bracket}
\{A,B\} \equiv \sum_n \frac{\partial A}{\partial q_n}
\frac{\partial B}{\partial p_n} - \frac{\partial B}{\partial q_n}
\frac{\partial A}{\partial p_n}
\end{equation}
for canonical positions $q_n$ and canonical momenta $p_n$. Then the condition (\ref{closure}) amounts to
\begin{equation}
\{Q_a,Q_b \} = f_{ab}^{~~c} Q_c + k\, \Delta_{ab}
\end{equation}
where the only restrictions are that $k\, \Delta_{ab}$ should have
vanishing Poisson brackets with the fields, and $\Delta_{ab}$ is
antisymmetric in its indices. This is because
\begin{equation}
\{f_{ab}^{~~c}Q_c +k\, \Delta_{ab},\phi \} = \{f_{ab}^{~~c}Q_c,\phi\} = f_{ab}^{~~c}\delta_c
\phi.
\end{equation}
Of course, $k\, \Delta_{ab}$ may, in particular, vanish. Thus, the
possibility for a central charge in the Poisson bracket algebra that does not act as a generator derives from an ambiguity in the Noether charges.

An example that illustrates this is two-dimensional Liouville theory.  The Hamiltonian reads \cite{Gervais}
\begin{equation}
H = \int \left\{\frac{1}{2}\pi^2 +
\frac{1}{2}\left(\partial_{\sigma}\phi\right)^2 + e^{\phi} -
2\partial_{\sigma}^2 \phi \right\} d\sigma,
\end{equation}
where the ranges of the coordinates are $\tau \in (-\infty,\infty)$, $\sigma \in [0, \pi)$. The canonical momentum is $\pi = \partial_{\tau}\phi$. The last term
$-2\partial_{\sigma}^2 \phi$ is a boundary term which is added to
make sure the variation of the Hamiltonian is well-defined. That is, together with the boundary conditions
\begin{eqnarray}
\partial_{\sigma}\phi &=& -\sqrt{2}~\rho ~e^{\phi/2} \ \ \ \mathrm{at} \ \sigma = 0, \nonumber \\
\partial_{\sigma}\phi &=& \sqrt{2}~\rho ~e^{\phi/2} \ \ \ \mathrm{at} \ \sigma = \pi,
\end{eqnarray}
where $\rho$ is a scale parameter, the variation of the action with respect to $\phi(\tau, \sigma)$ vanishes neatly at the endpoints of $\sigma$. The corresponding equation of motion is the Liouville equation
\begin{equation}
\phi_{\tau\tau}- \phi_{\sigma\sigma} + e^{\phi} = 0.
\end{equation}
This theory is conformally invariant. The Noether charges can be expanded according to
\begin{eqnarray}
L_0 &=& -\int \left\{ (\frac{1}{2}\pi^2 + \frac{1}{2}\left(\partial_{\sigma}\phi \right)^2 + e^{\phi} - 2\partial_{\sigma}^2 \phi ) \right\} d\sigma -2\pi \nonumber \\
L_m &=& -\int \bigg\{ (\frac{1}{2}\pi^2 + \frac{1}{2}\left(\partial_{\sigma}\phi \right)^2 + e^{\phi} - 2\partial_{\sigma}^2 \phi ) \cos{m\sigma} \nonumber \\
&&\ \ \ \ \ \ \ \ \ \ \ \ \ \ \ \ \ \ \ \ \ \ \ \ \  +~ i(\pi \partial_{\sigma}\phi - 2 \partial_{\sigma}\pi )\sin{m\sigma}\bigg\}d\sigma,
\end{eqnarray}
and their Poisson bracket algebra is the Virasoro algebra
\begin{equation}
\{L_m, L_n \} = i (m-n)L_{m+n} - 4\pi i (m^3-m)\delta_{m+n,0}.
\end{equation}
The central extension is claimed \cite{Gervais} to be a result of the presence of a boundary term in the Hamiltonian. As opposed to the previous example, the central charge of this theory is not a generator on the classical configuration space. Chapter 6 will reveal that two copies of precisely such an algebra are encountered when describing the asymptotic symmetries of a class of AdS$_3$-like metrics.

\chapter{Constraints: An Example from Maxwell Theory}

General relativity is sometimes regarded as the gauge theory of diffeomorphisms, where the role of the gauge field is played by the metric. Gauge invariance is reflected in the Hamiltonian formalism by relations between the canonical
variables, so-called \emph{constraints}. If the canonical positions are $q_n$ and the canonical momenta $p_n$, they can be expressed by $\phi_m(q_n,p_n) = 0$. Some of the $\phi_m$ generate gauge transformations. Constraints also arise in the Hamiltonian formulation of general relativity. In order to get an intuitive feeling, we first discuss the relatively simple example of free Maxwell theory, before turning to the more complicated case of general relativity in the next chapter.

As mentioned in section 2.2, the free Maxwell Lagrangian is
\begin{equation}\label{f}
L(A_{\mu}, \partial_0 A_{\mu}) = -\frac{1}{4} \int F^{\mu\nu}F_{\mu\nu} ~d^3 x ,
\end{equation}
with the field strength tensor $F_{\mu\nu} = \partial_{\mu}A_{\nu} - \partial_{\nu}A_{\mu}$ and $A_{\mu}$ the photon field. The index $0$ stands for the time parameter that needs to be integrated over in order to obtain the full action. For simplicity, we consider the theory in flat Minkowski space-time. The components of the field strength tensor are
\begin{equation}
F_{0i} = E_i,\ \ \  F_{ij} = -\epsilon_{ijk}B^{k},
\end{equation}
where $\epsilon_{ijk}$ is the fully antisymmetric Levi-Civita
symbol, $E_i$ is the electric field and $B_i$ the magnetic field.
Conversely, we then have $B^{k} = \epsilon^{ijk}\partial_i A_j$,
or in vector notation $\vec B = \vec \nabla \times \vec A$.

The equations of motion
\begin{equation}
\partial_{\mu}\frac{\delta L}{\delta(\partial_{\mu}A_{\nu})} = \frac{\delta L}{\delta A_{\nu}}
\end{equation}
read
\begin{equation}
\partial_{\mu}F^{\mu\nu} = 0.
\end{equation}
The remaining Maxwell equations follow from the Bianchi identity
\begin{equation}
\partial_{\mu}F_{\nu\rho} + \partial_{\rho}F_{\mu\nu} + \partial_{\nu}F_{\rho\mu} = 0.
\end{equation}

According to the rules of field theory, the canonical `position' becomes the field $A_{\mu}$, and the corresponding `momentum' is
\begin{equation}
\pi^{\mu} \equiv \frac{\delta L}{\delta(\partial_{0}A_{\mu})} = F^{\mu 0}.
\end{equation}
We see that $\pi^i$ is just the electric field. It is already clear that the canonical momentum $\pi^{0}$ vanishes due to the antisymmetry of $F^{\mu\nu}$. In other words, $\pi^{0}$ is one of the constraints of the theory, on a par with the identity $\vec \nabla^{.} \vec B = 0$ (telling us there are no magnetic monopoles). Because we can immediately see it vanishes, it is called a \emph{primary} constraint. 
So-called \emph{secondary} constraints may follow from the requirement that the primary constraints be consistent under time evolution. One can repeat this procedure until no new constraints arise.

Performing the Legendre transformation from the Lagrangian to the Hamiltonian yields a Hamiltonian that is only defined up to the addition of constraints, the \emph{total} Hamiltonian ($\mathcal{L}$ denotes the Lagrangian density)
\begin{eqnarray}\label{hamiltonian}
H_T(A_{\mu}, \pi^{\mu}) &=& \int \left\{ (\dot A_{\mu})\pi^{\mu} - \mathcal{L}\right\} ~d^3 x \nonumber \\
&=& \int \left\{ \dot A_0 \pi^0 + \dot A_i F^{i0} - \frac{1}{2}F^{i0}F_{i0} + \frac{1}{4}F^{ij}F_{ij} \right\} ~d^3 x \nonumber \\
&=& \int \left\{ \frac{1}{2}\pi^i \pi_i + \frac{1}{4}F^{ij}F_{ij} - A_0 (\partial_i \pi^i) + \dot A_0 \pi^0 \right\} ~d^3 x.
\end{eqnarray}
Notice that the transformation from the Lagrangian to the Hamiltonian formulation is invertible only if we treat $\dot A_0$ as a Lagrange multiplier for the constraint $\pi^0 = 0$. The Hamiltonian (\ref{hamiltonian}) further differs from the more familiar
\begin{equation}
H = \int \frac{1}{2}(\vec E^2 + \vec B^2)~d^3 x
\end{equation}
by the term -$A_0(\vec \nabla^{.}\vec E)$. We can thus expect $-\partial_i \pi^i$ to be another constraint, representing Gauss's law. We will now derive how this comes about on formal grounds.

As mentioned above, $\pi^0 = 0$ should be preserved in time in order for this to be a consistent constraint. When calculating the time derivative of $\pi^0$ we cannot simply set it to zero from the outset. The time evolution belonging to (\ref{hamiltonian}) is
\begin{equation}\label{time evolution}
\dot f = \bigl\{f, H\bigr\}
\end{equation}
for any function $f(A_{\mu}, \pi^{\mu})$ that does not depend explicitly on time. Choosing $f =
\pi^0$, we have
\begin{eqnarray}\label{consistency}
\left\{ \pi^0(\vec x), H \right\} &=& \int \bigg( \Big\{\pi^0(\vec x), \frac{1}{2}\pi^i\pi_i(\vec x') \Big\} + \Big\{\pi^0(\vec x),\frac{1}{4}F^{ij}F_{ij}(\vec x') \Big\} \nonumber \\
&& - \left\{\pi^0(\vec x), A_0(\partial_i\pi^i)(\vec x')\right\} + \left\{\pi^0(\vec x), \dot A_0\pi^0(\vec x') \right\} \bigg) d\vec x' \nonumber \\
&=& \int \!\!\int \bigg\{ \frac{\delta A_0(\partial_i \pi^i)(\vec x')}{\delta A_{\rho}(\vec x'')} \frac{\delta \pi^0(\vec x)}{\delta \pi^{\rho}(\vec x'')}  \bigg\} d\vec x' d\vec x'' \nonumber \\
&=& -\partial_i \pi^i (\vec x) \nonumber \\
&\equiv& 0.
\end{eqnarray}
We have thus found a secondary constraint which is recognized as the source-free Gauss law $ \vec \nabla^{.} \vec E = 0$. No further constraints arise, since the bracket of $\partial_i \pi^i$ with the Hamiltonian is zero.

The surface defined by $\pi^0 = \partial_i\pi^i = 0$ is a smooth submanifold of the configuration (or phase) space called the constraint surface. In particular, the physical content of a function on phase space is represented by the value of the function on this surface. We call functions that may differ off the constraint surface, but coincide on it, \emph{weakly} equal. Observe that there may be many physically distinct solutions satisfying the constraints. Each of these can be evolved with the Hamiltonian according to (\ref{time evolution}). There exists a similar initial value formulation of general relativity.

In the derivation of both equations of motion and constraints, no restriction whatsoever has been placed on the Lagrange multipliers $A_0$ and $\dot A_0$. Thus, the time evolution (\ref{time evolution}) contains arbitrary functions. Since, given a Hamiltonian, we assume the physical content of a state to fully determine its physical content at another moment in time, the difference resulting from a difference in $A_0$ and $\dot A_0$ should be unobservable; in other words, it should be a gauge transformation. This is in agreement with the so-called Dirac conjecture, which states that all \emph{first-class} constraints - constraints which have vanishing Poisson brackets with all other constraints - are generators of gauge transformations.

To see what the Gauss constraint generates, it is convenient to integrate with a parameter $\lambda(\vec x')$:
\begin{eqnarray}\label{gauss constraint transformation}
&& \{A_k(\vec x),\int \partial_i \pi^i ( \vec x') \lambda( \vec x') d^3 x' \} \nonumber \\
&=& \int \left\{ \frac{\delta A_k(\vec x)}{\delta A_j(\vec x'')}\frac{\delta(\int \partial_i \pi^i (\vec x') \lambda(\vec x') d^3 x')}{\delta \pi^j(\vec x'')} - \frac{\delta(\int \partial_i \pi^i (\vec x') \lambda(\vec x') d^3 x')}{\delta A_j(\vec x'')}\frac{\delta A_k(\vec x)}{\delta \pi^j(\vec x'')} \right\} d^3 x'' \nonumber \\
&=& \int \left\{ \delta^k_{~j} ~\delta(\vec x - \vec x'') \frac{\delta(-\int \pi^i (\vec x') \partial_i \lambda(\vec x') d^3 x')}{\delta \pi^j(\vec x'')} \right\} d^3 x'' \nonumber \\
&=& \partial_k \lambda(\vec x).
\end{eqnarray}
This is recognized as the $U(1)$ gauge transformation
\begin{equation}\label{uone}
\delta A_i = \partial_i\lambda,
\end{equation}
and we see that this first-class constraint indeed generates gauge transformations. All in all, the arbitrariness of $A_0$ together with the gauge freedom (\ref{uone}) tells us that only two polarizations of the photon have physical significance. The other two components of the vector potential $A_{\mu}$ can be consistently gauged away. A common gauge choice is the Coulomb or radiation gauge $A_0 = 0$, $\partial_iA^i = 0$, which leaves no room for the transformation (\ref{gauss constraint transformation}). Instead of $A_0 = 0$, we may also simply exclude $A_0$ and $\pi^0$ from the phase space altogether.

In the above derivation we have silently stepped over an issue concerning the canonical bracket itself. In the present example we have been fortunate enough to have only first-class constraints. In general, however, constraints may arise that do not have weakly vanishing Poisson brackets with the other constraints. These are called \emph{second-class} constraints. If the conservation in time of some constraints places restrictions on the Lagrange multipliers, this is indicative of the presence of second-class constraints. Whereas the freedom in the Lagrange multipliers of the first-class constraints led to an arbitrary contribution to the equations of motion, the multipliers of second-class constraints are not arbitrary, and the latter do not (in general) generate gauge transformations. Now the fundamental point to be made about first-class versus second-class constraints is that (after deriving the full set of constraints) those that are first-class can be set to zero either before or after evaluating the Poisson bracket, as they do not alter the physical content of a state. This is not the case for second-class constraints. The problem is solved, however, by introducing the Dirac bracket
\begin{equation}\label{Dirac}
\{A, B \}^{*} = \{A, B \} - \{A, \chi_{\alpha} \} \{\chi_{\alpha}, \chi_{\beta} \}^{-1}  \{\chi_{\beta}, B \},
\end{equation}
where $\chi_{\alpha}$ are the second-class constraints, and they are assumed to be independent. This can always be achieved. The bracket (\ref{Dirac}) enables one to set $\chi_{\alpha} = 0$ either before or after its evaluation, since the Dirac bracket of second-class constraints with other functions on phase space vanishes weakly. Moreover, if either $A$ or $B$ is first-class, the bracket reduces weakly to the Poisson bracket. The Hamiltonian, in particular, is first-class by construction, and Hamiltonian evolution is unaffected by the replacement of Poisson brackets by Dirac brackets.

\chapter{Hamiltonian Formulation of General Relativity}

In classical mechanics, the Hamiltonian generates time
translations of the canonical variables through the Poisson
bracket. Similarly, the diffeomorphisms of general relativity are generated by functions that differ from the Hamiltonian only through a simple substitution of vectors.

The asymptotically anti-de Sitter space-times have a boundary at spatial infinity, which gives a contribution to the Hamiltonian referred to as the \emph{surface charge}. In the present chapter, we derive this surface charge starting from the Lagrangian formulation of general relativity with a cosmological constant. Section 4.1 first introduces the notions of ADM decomposition and extrinsic curvature. The former is useful for defining the Hamiltonian; the latter for deriving the surface term in the action, which is done in section 4.2. Section 4.3 finally brings us to the surface charge, which is shown to become the generator of diffeomorphisms, as the bulk term of the Hamiltonian vanishes weakly. Another form of the surface charge, coming from \cite{Brown-Henneaux}, is given.

\section{ADM Decomposition}
The Hamiltonian formulation of general relativity makes use of a
special decomposition of the metric, called the Arnowitt-Deser-Misner decomposition \cite{ADM} \mbox{(ADM for short)}, which brings out the
components parallel and orthogonal to hypersurfaces at constant
$t$. The decomposition is useful for separating the canonical variables into constraints and physically meaningful quantities, analogous to what happens in Maxwell theory.

Suppose we have a global time function $t(x^{\mu})$ and a vector
field $t^{\mu}$ obeying $t^{\mu}\partial_{\mu}t = 1$. In
particular, we can introduce a coordinate system $(t, x^a)$ with
as time coordinate the global time function $t$. The fact that $t$
is global implies that we can foliate our space-time into
hypersurfaces at constant $t$, and $t^{\mu}$ generates the flow
from the initial spacelike hypersurface $\Sigma_0$ to $\Sigma_t$.
If we choose $t$ to be one of the coordinates of a frame, we
simply have $t^{\mu} = \left(\frac{\partial}{\partial
t}\right)^{\mu}$, since in this case $\frac{\partial
x^{\mu}}{\partial t} \frac{\partial t}{\partial x^{\mu}} = 1$.
From now on we will treat $t$ as a coordinate. We decompose the
vector $t^{\mu}$ in terms of the vector $n^{\mu}$ normal to
surfaces at constant $t$ and an additional vector $N^{\mu}$:
\begin{equation}\label{flow of time}
t^{\mu} \equiv N n^{\mu} + N^{\mu}.
\end{equation}
$N$ is the so-called lapse function and $N^{\mu}$ the shift
vector. The decomposition (\ref{flow of time}) shows that the
lapse function is the normal component and the shift vector is the
parallel component of $t^{\mu}$ to
surfaces at constant $t$.

We can use lapse and shift to obtain the ADM form of the metric
starting from the following metric adapted to the hypersurface,
\begin{equation}\label{ADM}
ds^2 = -(dx^{\perp})^2 + \omega_{ab}(x^{\perp}, x^{a})dx^{a}dx^{b},
\end{equation}
where $a$ and $b$ run from $1$ to $d-1$, and the coordinates
$x^{\perp}, x^a$ are independent. This metric is adapted to the
family of hypersurfaces at constant $t$ in the sense that the
coordinates are so defined that the component of $n^{\mu}$ in the
$x^{\perp}$-direction becomes $1$, while the components in the
$x^a$-directions vanish. Since we have $t^{\mu} =
\left(\frac{\partial}{\partial t}\right)^{\mu}$, the lapse and
shift become
\begin{eqnarray}
N &=&
\frac{\partial
x^{\perp}}{\partial t} \nonumber \\
N^{a} &=&
\frac{\partial x^{a}}{\partial t}.
\end{eqnarray}
in the hypersurface-adapted frame. We can use this to turn to
another frame $(t(x^{\perp},x^a), x^a)$. The tensor transformation rule (denoting the
coordinates appearing in (\ref{ADM}) collectively by $\hat
x^{\rho} = (x^{\perp}, x^{a})$, using a Greek index, and writing
$x^{\mu} = (t, x^a)$)
\begin{equation}
g_{\mu\nu}(t, x^{a}) = \frac{\partial \hat x^{\rho}}{\partial x^{\mu}} \frac{\partial \hat x^{\sigma}}{\partial x^{\nu}} \hat g_{\rho\sigma}(x^{\perp}, x^{a})
\end{equation}
then tells us that
\begin{eqnarray}
g_{tt} = -N^2 + \omega_{ab}N^{a}N^{b},\ \ \ g_{ta} =
\omega_{ab}N^{b},\ \ \ g_{ab} = \omega_{ab},
\end{eqnarray}
yielding the following line element in the coordinates $t, x^{a}$:
\begin{equation}
ds^2 = -N^2 dt^2 + \omega_{ab}(dx^{a} + N^{a}dt)(dx^{b} +
N^{b}dt).
\end{equation}
This form of the metric is called the ADM decomposition.

Note that, due to the possible dependence of the metric on
$x^{\perp}$ and $t$, neither are necessarily global coordinates.
This will only be the case for specific solutions with Killing
vectors $\frac{\partial}{\partial x^{\perp}}$ or
$\frac{\partial}{\partial t}$, respectively. We called $t$ a
\emph{global} time function only because the space-time can be
foliated into hypersurfaces at constant $t$, which is indicative
of a product topology.

For space-times with the appropriate product topology, the
decomposition can also be done with respect to other foliations.

In the following, we will also need the notion of an induced metric.
The line element induced on hypersurfaces of constant $t$ is in the present example
\begin{equation}
ds^2 = \omega_{ab}N^a N^b dt^2 + 2\omega_{ab}N^a dtdx^b + \omega_{ab}dx^a dx^b
\end{equation}
This follows from the more general definition of an induced metric,
\begin{equation}
\gamma_{\mu\nu} = g_{\mu\nu} \pm n_{\mu}n_{\nu},
\end{equation}
with $n^{\mu}$ the unit normal to a spacelike or timelike
hypersurface, respectively. In order for volume elements to keep the right orientation, $n^{\mu}$ should be inward pointing if it is timelike and outward pointing if it is spacelike \cite{Wald}. If part of the boundary
is, for instance, a timelike hypersurface at constant radius, then for this part the normal is
\begin{eqnarray}
n^{\mu} &=& \frac{1}{\sqrt{g^{rr}}}g^{\mu(r)},\ \ \ n_{\mu} =
\frac{1}{\sqrt{g^{rr}}}\delta^{(r)}_{~\mu}, \nonumber \\
 n_{\mu}n^{\mu} &=& 1,
\end{eqnarray}
and the radial contravariant components of the induced metric
vanish. We keep $d$-valued indices, however, because as we saw the covariant components are generally nonvanishing.

Indices
of tensors on the boundary, including the induced metric, are still to be raised and lowered with the full space-time metric $g_{\mu\nu}$. However, $\gamma^{\mu\nu}$ will not be the inverse of $\gamma_{\mu\nu}$, which is noninvertible since we have removed the components in the normal direction. Note that we have $\sqrt{-g} = N\sqrt{\pm \gamma}$, with $N$ the appropriate lapse function. Moreover, $\gamma_{\mu\nu}n^{\mu} = 0$.

A final notion we introduce here is that of the extrinsic
curvature of a hypersurface. Its definition is given in terms of
the induced metric $\gamma_{\mu\nu}$ and the normal vector
$n^{\mu}$ as
\begin{equation}\label{extrinsic curvature}
\Theta_{\mu\nu} = \gamma^{\rho}_{~\mu}\gamma^{\sigma}_{~\nu}
D_{\rho}n_{\sigma},
\end{equation}
where the covariant derivative contains the full $d$-dimensional connection. The $\gamma^{\rho}_{~\mu} = g^{\rho\nu}\gamma_{\nu\mu} = \delta^{\rho}_{~\mu} \pm n^{\rho}n_{\mu}$ work as projection operators of
tensors onto the hypersurface. If $n^{\mu}$ is a geodesic normal, the
projection in (\ref{extrinsic curvature}) is unnecessary. We can write (\ref{extrinsic
curvature}) in a more intuitive form by making use of the fact
that for $n^{\mu}$ to be hypersurface orthogonal is equivalent to
(square brackets denote antisymmetrization over all indices)
\begin{equation}\label{Frobenius}
n_{[\mu}D_{\nu}n_{\rho]} = 0,
\end{equation}
a result following from Frobenius's theorem on integral
submanifolds. Suppose we are trying to find integral surfaces
(i.e. surfaces that are everywhere parallel to the original
surface, and span the space) of an $m$-dimensional surface with
$d-m$ linearly independent normal covectors $n^{(a)}_{\nu}, ~a =
1, \ldots, d-m$ living in the cotangent space of a $d$-dimensional
manifold. The theorem states that in order for this surface to be
integrable, such $n_{\nu}$ should obey $D_{[\mu}n_{\nu]} = \sum_b
\eta^{(b)}_{[\mu}v^{(b)}_{\nu]}$, where the $\eta_{\mu}^{(b)}$, $b
= 1, \ldots, d-m$, lie in the cotangent space orthogonal to the
surface, and $v_{\nu}$ are some covectors. If, in particular, $m = d-1$, we obtain the result
(\ref{Frobenius}) by substituting $\eta_{\mu} = n_{\mu}$.

Contracting (\ref{Frobenius}) with $n^{\nu}$, we have
\begin{equation}\label{contracted}
D_{\rho}n_{\mu} + n^{\nu}n_{\mu}D_{\nu}n_{\rho} = D_{\mu}n_{\rho} + n^{\nu}n_{\rho}D_{\nu}n_{\mu},
\end{equation}
since $n^{\nu}D_{\mu}n_{\nu} = \frac{1}{2}D_{\mu}(n^{\nu}n_{\nu}) = 0$. Furthermore,
\begin{eqnarray}
n^{\nu}n_{\mu}D_{\nu}n_{\rho} &=& \mp g^{\nu}_{~\mu}D_{\nu}n_{\rho} \pm \gamma^{\nu}_{~\mu}D_{\nu}n_{\rho} \nonumber \\
&=& \mp D_{\mu}n_{\rho} \pm \gamma^{\nu}_{~\mu}D_{\nu}n_{\rho}.
\end{eqnarray}
If the lower signs apply, substituting this in (\ref{contracted}) gives
\begin{equation}
2D_{\mu}n_{\rho} + 2D_{\rho}n_{\mu} = \gamma^{\nu}_{~\rho}D_{\nu}n_{\mu} - \gamma^{\nu}_{~\mu}D_{\nu}n_{\rho}.
\end{equation}
Since one side is symmetric under $\rho \leftrightarrow \mu$, while the other side is antisymmetric under this exchange, both sides are equal to zero.
The upper signs immediately give
\begin{equation}
\gamma^{\nu}_{~\mu}D_{\nu}n_{\rho} = \gamma^{\nu}_{~\rho}D_{\nu}n_{\mu},
\end{equation}
and in both cases the expression is symmetric in $\mu$ and $\rho$. At the same time, using $\gamma^{\alpha}_{~\mu} = g^{\alpha}_{~\mu} \pm n^{\alpha}n_{\mu}$, it can be verified that (\ref{extrinsic curvature}) can also be written simply as $\gamma^{\alpha}_{~\mu}D_{\alpha}n_{\nu}$. Thus, through a relatively complicated procedure we have verified that the extrinsic curvature is actually symmetric in its indices, and we can now write it as the Lie derivative of $\gamma_{\mu\nu}$ in the normal direction:
\begin{equation}
\Theta_{\mu\nu} = \frac{1}{2}\mathcal{L}_{n} \gamma_{\mu\nu}
\end{equation}
This more intuitive form of $\Theta_{\mu\nu}$ tells us that the extrinsic curvature measures the rate of change of the induced metric when
moving off the hypersurface. $\Theta_{\mu\nu}$ will turn out to be a useful quantity in defining the
Einstein-Hilbert action for space-times with boundary, as is discussed in the next section.

\section{Surface Term in the Action}
In order to pass to the Hamiltonian formulation, we first need to know
which action to use. For a space-time $M$ without boundary,
Einstein's equations can be derived from the action
\begin{equation}\label{EH action}
S = \frac{1}{16 \pi G} \int_{M} \sqrt{-g} ~(R - 2\Lambda)~d^{d}x
\end{equation}
by varying with respect to the metric $g_{\mu\nu}$. Here, G is
Newton's constant in $d$ dimensions, $R$ the Ricci scalar
curvature, and $\Lambda$ the cosmological constant. We do not include matter terms, since we will be interested in vacuum solutions.

For a space-time with boundary, the action (\ref{EH action}) also
gives Einstein's equation as long as boundary terms are ignored in
the variational principle. However, we wish to use a different
variational principle, in which we only set $\delta g_{\mu\nu} =
0$ at the boundary, but do not throw away surface terms in the
variation of the action. These can still arise because we do not
demand the derivative of $g_{\mu\nu}$ to vanish at the boundary.

To determine the boundary term in the action, we will write out
the general variation of the Einstein-Hilbert action, and see
which term is needed to cancel the unwanted surface term. It is
useful to write the Riemann curvature tensor as
\begin{equation}
R^{\sigma}_{~\mu\nu\rho} = \partial_{\nu}\Gamma^{\sigma}_{~\mu\rho} - \partial_{\rho}\Gamma^{\sigma}_{~\mu\nu} + \Gamma^{\sigma}_{~\nu\kappa}\Gamma^{\kappa}_{~\mu\rho} - \Gamma^{\sigma}_{~\rho\kappa}\Gamma^{\kappa}_{~\mu\nu},
\end{equation}
so that we can recognize its variation,
\begin{equation}
 \partial_{\nu}\delta\Gamma^{\sigma}_{~\mu\rho} - \partial_{\rho}\delta\Gamma^{\sigma}_{~\mu\nu} + \Gamma^{\kappa}_{~\mu\rho}\delta\Gamma^{\sigma}_{~\nu\kappa} + \Gamma^{\sigma}_{~\nu\kappa}\delta\Gamma^{\kappa}_{~\mu\rho} - \Gamma^{\sigma}_{~\rho\kappa}\delta\Gamma^{\kappa}_{~\mu\nu} - \Gamma^{\kappa}_{~\mu\nu}\delta\Gamma^{\sigma}_{~\rho\kappa},
\end{equation}
as
\begin{equation}
D_{\nu}\delta\Gamma^{\sigma}_{~\mu\rho} - D_{\rho}\delta\Gamma^{\sigma}_{~\mu\nu}.
\end{equation}
The variation of (\ref{EH action}) thus becomes
\begin{eqnarray}
\delta S &=&\frac{1}{16 \pi G} \int_{M} \delta(R^{\sigma}_{~\mu\nu\rho}\sqrt{-g}\delta^{\nu}_{~\sigma}g^{\mu\rho}) - 2\Lambda \delta(\sqrt{-g}) ~d^d x \nonumber \\
&=& \frac{1}{16 \pi G} \int_{M} \sqrt{-g} \{ \delta^{\nu}_{~\sigma}g^{\mu\rho}(D_{\nu}\delta\Gamma^{\sigma}_{~\mu\rho} - D_{\rho}\delta\Gamma^{\sigma}_{~\mu\nu}) \nonumber \\
&+& \delta^{\nu}_{~\sigma} R^{\sigma}_{~\mu\nu\rho} \delta g^{\mu\rho}
- \frac{1}{2} \delta^{\nu}_{~\sigma}g^{\mu\rho}R^{\sigma}_{~\mu\nu\rho} g_{\alpha\beta}\delta g^{\alpha\beta}
+ \Lambda g_{\alpha\beta}\delta g^{\alpha\beta} \} ~d^d x
\end{eqnarray}
The term on the second line is easily recognized as a total derivative, and applying Gauss's law
\begin{equation}
\int_M \sqrt{-g} ~D_{\alpha}v^{\alpha} ~d^d x = \int_{\partial M} \sqrt{\pm\gamma} ~n_{\alpha}v^{\alpha} ~d^{d-1}x
\end{equation}
where $\gamma$ is the determinant of the metric induced on the
boundary, and $n_{\alpha}$ a unit normal to this boundary, now
yields
\begin{eqnarray}\label{total variation}
\delta S = \frac{1}{16 \pi G} \int_{M} \sqrt{-g}~(R_{\mu\nu} - \frac{1}{2}R g_{\mu\nu} + \Lambda g_{\mu\nu})\delta g^{\mu\nu}~d^d x \nonumber \\
+ \frac{1}{16 \pi G} \int_{\partial M} \sqrt{\pm \gamma} ~(g^{\mu\rho}\delta\Gamma^{\nu}_{~\mu\rho} - g^{\mu\nu}\delta\Gamma^{\rho}_{~\mu\rho})n_{\nu}~d^{d-1}x.
\end{eqnarray}
The first term alone would give Einstein's equations
\begin{equation}
R_{\mu\nu} - \frac{1}{2}R g_{\mu\nu} + \Lambda g_{\mu\nu} = 0.
\end{equation}
The remaining term is further reduced using
\begin{equation}\label{gamma variation}
\delta \Gamma^{\nu}_{~\mu\rho} = \frac{1}{2}g^{\nu\sigma}(D_{\mu}\delta g_{\sigma\rho} + D_{\rho}\delta g_{\mu\sigma} - D_{\sigma}\delta g_{\mu\rho}),
\end{equation}
which tells us that
\begin{eqnarray}
&& g^{\mu\rho}\delta\Gamma^{\nu}_{~\mu\rho} - g^{\mu\nu}\delta\Gamma^{\rho}_{~\mu\rho} \nonumber \\
&=& \frac{1}{2}(g^{\mu\rho}g^{\nu\sigma} - g^{\mu\nu}g^{\rho\sigma})(D_{\mu}\delta g_{\rho\sigma} + D_{\rho}\delta g_{\mu\sigma} - D_{\sigma}\delta g_{\mu\rho}) \nonumber \\
&=& g^{\mu\nu}g^{\rho\sigma}(D_{\sigma}\delta g_{\mu\rho} - D_{\mu}\delta g_{\rho\sigma}),
\end{eqnarray}
since $D_{\rho}\delta g_{\mu\sigma}$ is symmetric under $\mu \leftrightarrow \sigma$, while $D_{\mu}\delta g_{\rho\sigma} - D_{\sigma}\delta g_{\mu\rho}$ is antisymmetric under this exchange, as is the prefactor.
This leads to
\begin{eqnarray}
n_{\nu}(g^{\mu\rho}\delta\Gamma^{\nu}_{~\mu\rho} - g^{\mu\nu}\delta\Gamma^{\rho}_{~\mu\rho}) &=& n^{\mu}g^{\rho\sigma}(D_{\sigma}\delta g_{\mu\rho} - D_{\mu}\delta g_{\rho\sigma}) \nonumber \\
&=& n^{\mu}\gamma^{\rho\sigma}(D_{\sigma}\delta g_{\mu\rho} - D_{\mu}\delta g_{\rho\sigma}) \nonumber \\
&=& -n^{\mu}\gamma^{\rho\sigma}D_{\mu}\delta g_{\rho\sigma}.
\end{eqnarray}
To get the third line we again used the antisymmetry of
$D_{\sigma}\delta g_{\mu\rho} - D_{\mu}\delta g_{\rho\sigma}$
under $\mu \leftrightarrow \sigma$, so that
$n^{\mu}n^{\rho}n^{\sigma}(D_{\sigma}\delta g_{\mu\rho} -
D_{\mu}\delta g_{\rho\sigma}) = 0$. The last equality follows from
the fact that $\gamma^{\rho\sigma}D_{\sigma}$ is the covariant
derivative along the boundary, where $\delta g_{\mu\rho}$ vanishes
everywhere. We have now written the boundary term in a form which
allows us to recognize it as the variation of a tensor living on
the boundary. Recall that we had
\begin{eqnarray}
\Theta_{\mu\nu} &=& \gamma^{\rho}_{~\mu}\gamma^{\sigma}_{~\nu}
D_{\rho}n_{\sigma}
\nonumber \\
&=& \frac{1}{2}\mathcal{L}_{n} \gamma_{\mu\nu}.
\end{eqnarray}
The trace of the extrinsic curvature of the boundary is then given by
\begin{equation}
\Theta = \gamma^{\rho\sigma}D_{\rho}n_{\sigma} =
\gamma^{\rho\sigma}\partial_{\rho}n_{\sigma} -
\gamma^{\rho\sigma}\Gamma^{\mu}_{~\rho\sigma}n_{\mu}.
\end{equation}
From the boundary condition $\delta g_{\mu\nu} =
0$ and the fact that a variation of $\gamma_{\mu\nu}$ cannot
cancel a variation of $n_{\mu}$, it follows that $\delta n_{\mu} = \delta \gamma_{\mu\nu} = 0$ at $\partial M$.
Hence we have
\begin{equation}
\delta \Theta = -\gamma^{\rho\sigma}\delta \Gamma^{\mu}_{~\rho\sigma}n_{\mu}
\end{equation}
Using (\ref{gamma variation}) this reduces to
\begin{equation}
\delta \Theta = \frac{1}{2} n^{\mu}\gamma^{\rho\sigma}D_{\mu}\delta g_{\rho\sigma},
\end{equation}
or $-\frac{1}{2}$ times the surface term in the variation of
$\int_M \sqrt{-g}~(R -2\Lambda) d^d x$. This finally motivates
the use of the adapted Einstein-Hilbert action
\begin{equation}\label{with surface term}
S = \frac{1}{16 \pi G} \int_{M} \sqrt{-g} ~(R - 2\Lambda) ~d^{d}x
+ \frac{1}{8 \pi G} \int_{\partial M} \sqrt{\pm \gamma} ~\Theta
~d^{d-1}x.
\end{equation}
This action gives the Einstein equations of motion both for
space-times without boundary and for space-times with boundary
under the condition that $\delta g_{\mu\nu} = 0$ at $\partial M$,
and taking into account surface terms. The action (\ref{with
surface term}) is the one we will be working with.

\section{Hamiltonian Formalism}

The ADM decomposition now comes in handy, because it allows us to write the action (\ref{with surface term}) in terms of the metric $h_{\mu\nu}$ induced on surfaces of constant $t$ and only its first time derivative $\dot h_{\mu\nu}$, since these two quantities contain all the physical information necessary to describe the dynamics. This in turn allows us to define a canonical momentum
\begin{equation}
\pi^{\mu\nu} \equiv \frac{\delta S}{\delta \dot h_{\mu\nu}}.
\end{equation}
Passage to the Hamiltonian formulation will then be
achieved through the definition
\begin{equation}
S \equiv \int^{t_f}_{t_i} \left(\int_{\Sigma_{t}} \pi^{\mu\nu}\dot
h_{\mu\nu} - H ~d^{d-1}x\right) ~dt.
\end{equation}
This shows that properly defining the Hamiltonian is
only possible if the space-time has a global time coordinate, and
can be foliated into surfaces $\Sigma_{t}$ at constant $t$:
\begin{equation}
M = [t_{i}, t_{f}] \times \Sigma_{t}.
\end{equation}
Assuming our space-time knows of such a thing as spatial infinity
(and the spatial part is not, for instance, a torus or the surface of a sphere), the boundary $\partial M$
consists of initial and final spacelike boundaries, and a timelike
boundary at spatial infinity:
\begin{equation}
\partial M = \Sigma_{t_{i}} + \Sigma_{t_{f}} + \Sigma^{\infty}
\end{equation}
Thus, the action (\ref{with surface term}) becomes
\begin{eqnarray}\label{action written out}
S = \frac{1}{16 \pi G} \int_{M} \sqrt{-g} ~(R - 2\Lambda) ~d^{d}x
+ \frac{1}{8 \pi G} \int_{\Sigma_{\infty}} \sqrt{- h} ~\Theta
~d^{d-1}x \nonumber \\
+ \frac{1}{8 \pi G}\int_{\Sigma_{t_i}} \sqrt{\gamma}~K ~d^{d-1}x- \frac{1}{8\pi G} \int_{\Sigma_{t_f}}\sqrt{\gamma}~K~d^{d-1}x,
\end{eqnarray}
with $\gamma_{\mu\nu}$ the metric induced on the $\Sigma_{t_{i}}$ and $\Sigma_{t_f}$ boundaries, and $K$ the trace of their extrinsic curvature. Using the contracted Gauss-Codacci relation
\begin{equation}
^{(d-1)}\mathcal{R} = R_{\mu\nu\rho\sigma}h^{\mu\rho}h^{\nu\sigma} - \Theta^2 + \Theta_{\mu\nu}\Theta^{\mu\nu}
\end{equation}
to express the Ricci curvature scalar $^{(d-1)}\mathcal{R}$ belonging to $h_{\mu\nu}$ in terms of the full $d$-dimensional curvature $R_{\mu\nu\rho\sigma}$, the action (\ref{action written out}) can be written without a contribution from the spacelike boundaries\footnote{We assume the $\Sigma^{\infty}$ and $\Sigma_{t}$ hypersurfaces to be orthogonal at infinity, which may not always be true, but holds for the asymptotically anti-de Sitter space-times which we are interested in.}:
\begin{eqnarray}
S = \frac{1}{16 \pi G}\int_{\mathcal{M}} \sqrt{- g}~\left(^{(d-1)}\mathcal{R} - 2\Lambda - \Theta^2 + \Theta_{\mu\nu}\Theta^{\mu\nu} \right) ~d^d x \nonumber \\
+ \frac{1}{8 \pi G} \int_{\Sigma_{\infty}} \sqrt{-\gamma}~\left(K - u_{\mu}n^{\nu}D_{\nu}n^{\mu}\right)~d^{d-1}x.
\end{eqnarray}
Here, $u^{\mu}$ is the unit normal vector to the $\Sigma^{\infty}$ boundary, and $D_{\nu}$ is still the $d$-dimensional covariant derivative.
The surface $\Sigma^{\infty}$ can again be split up into
hypersurfaces $\Sigma^{\infty}_{t}$ at constant $t$ (in three
dimensions these are just closed lines). Using
\begin{equation}
\Theta_{\mu\nu} = \frac{1}{2N}(\dot h_{\mu\nu} - D_{\mu}N_{\nu} - D_{\nu}N_{\mu}),
\end{equation}
we find as the momentum conjugate to $h_{\mu\nu}$
\begin{eqnarray}
\pi^{\mu\nu} &\equiv& \frac{\delta S}{\delta \dot h_{\mu\nu}}
\nonumber \\
&=& \frac{1}{16 \pi G} \sqrt{h} ~(\Theta^{\mu\nu} - \Theta
h^{\mu\nu}).
\end{eqnarray}
(the purely spatial metric $h_{\mu\nu}$ will have a positive
determinant).

For a space-time with the appropriate product topology, the Hamiltonian then becomes \cite{Terashima}
\begin{eqnarray}\label{Hamiltonian}
H(N) = \int_{\Sigma_{t}} (N \mathcal{H} + N_{\mu}\mathcal{H}^{\mu}) ~d^{d-1}x \nonumber \\
- \int_{\Sigma^{\infty}_{t}} \sqrt{\sigma}~(\frac{1}{8 \pi G}N
\theta - \frac{2}{\sqrt{h}}s_{\mu}\pi^{\mu\nu}N_{\nu}) ~d^{d-2}x.
\end{eqnarray}
where $\sigma$ is the determinant of the induced metric on
$\Sigma^{\infty}_{t}$, $s^{\mu}$ the unit normal vector to this surface
(i.e., it is the normal to $\Sigma^{\infty}$ projected onto
$\Sigma_{t}$), and $\theta$ the trace of its extrinsic curvature.
$\mathcal{H}$ and $\mathcal{H}^{\mu}$ are the constraints of general relativity, and $N$ and $N_{\mu}$ act as their Lagrange multipliers. The constraints take the form
\begin{eqnarray}
\mathcal{H} &=& ^{(d-1)}\mathcal{R} - 2\Lambda + \Theta^2 - \Theta_{\mu\nu}\Theta^{\mu\nu} \nonumber \\
\mathcal{H}^{\mu} &=& \nabla_{\nu}\Theta^{\nu}_{~\mu} - \nabla_{\mu}\Theta,
\end{eqnarray}
where $\nabla_{\mu}$ is the covariant derivative belonging to $h_{\mu\nu}$. The dynamical equations now read
\begin{eqnarray}\label{spatial equations}
\dot h_{\mu\nu} = \{h_{\mu\nu},H(N)\}^*, \nonumber \\
\dot \pi^{\mu\nu} = \{\pi^{\mu\nu},H(N)\}^*,
\end{eqnarray}
for which the explicit expressions are given in Table 4.1, together with a summary of the analogy between source-free Maxwell theory and general relativity with a cosmological constant.
\begin{table}[!h]
\label{analogy}
\begin{tabular}{l|c|c}
& Constraints & Evolution Equations \\
\hline
& & \\
General &  $^{(d-1)}\mathcal{R} + \Theta^2 - \Theta^{\mu\nu}\Theta_{\mu\nu} = 2\Lambda$ & $\partial_t \gamma_{\mu\nu} = -2N\Theta_{\mu\nu} + \nabla_{\mu}N_{\nu} + \nabla_{\nu}N_{\mu}$  \\
Relativity & & \\
& $\nabla_{\nu}\Theta^{\nu}_{~\mu} - \nabla_{\mu}\Theta = 0$ & $(\partial_t - N^{\rho}\partial_{\rho})\Theta_{\mu\nu} =  - \nabla_{\mu}\nabla_{\nu}N - N\gamma_{\mu\nu}\Lambda $ \\
& & \ \ \ \ \ \ \ \ \ $+ N\left(^{(d-1)}\mathcal{R}_{\mu\nu} + \Theta \Theta_{\mu\nu} - 2 \Theta_{\mu\rho}\Theta^{\rho}_{~\nu}\right)$ \\
& & \ \ \ \ \ \ \ \ \ \ \ \ \ \ \ \ \ \ \ \ \ \ \ $+\Theta_{\rho\nu}\partial_{\mu}N^{\rho} + \Theta_{\mu\rho}\partial_{\nu}N^{\rho}$ \\
& & \\
\hline
& & \\
Maxwell & $\vec \nabla^{.}\vec E = 0$ & $\partial_t \vec E = \vec \nabla \times \vec B$  \\
Theory & $\vec \nabla^{.}\vec B = 0$ & $\partial_t \vec B = -\vec \nabla \times \vec E $ \\
& & \\
\hline
\end{tabular}
\caption{The analogy between Maxwell theory and general relativity.}
\end{table}

If initially $\mathcal{H} = \mathcal{H}^{\mu} = 0$, and the evolution equations hold everywhere, then the constraints will be satisfied at all times. Using the ADM decomposition, we have in effect made a change of variables from $(g_{\mu\nu}, \Gamma^{\mu}_{~\nu\rho})$ to $(h_{\mu\nu}, \pi^{\mu\nu}, \mathcal{H}, \mathcal{H}^{\mu}, N, N^{\mu})$, separating physically meaningful quantities from constraints.

Since $N_{\mu}$ effectively has only $d-1$ components, it is seen that the extended variational principle reduces the degrees of freedom by $d$. Starting out with $\frac{d(d-1)}{2}$ components of the symmetric, $(d-1)$-dimensional tensor $h_{\mu\nu}$, the graviton then has $\frac{d(d-3)}{2}$ physical polarizations, or $2$ in $d=4$. Also, pure three-dimensional gravity has no dynamical degrees of freedom at all. 

Something else that comes to our attention is that the Hamiltonian of general relativity is weakly zero if boundary terms are ignored. This is in fact characteristic of generally covariant systems for which the canonical variables transform as scalars under time reparametrizations. This is easily seen \cite{Teitelboim} from
\begin{equation}\label{weakly zero}
S = \int (p_{\mu}\dot q^{\mu} - \lambda^m \phi_m - H_{\mathrm{rest}}(p, q)) dt
\end{equation}
where $\phi_m$ are the first- and second-class constraints, $\lambda^m$ their Lagrange multipliers, and $H_{\mathrm{rest}}(p, q)$ represents any part of the Hamiltonian that does not vanish weakly. The action (\ref{weakly zero}) is invariant under the transformation $t \to t - \epsilon(t)$ provided the various variables transform as
\begin{eqnarray}
\delta q &=& \dot q \epsilon \nonumber \\
\delta p &=& \dot p \epsilon \nonumber \\
\delta \lambda^m &=& \dot \lambda^m \epsilon + \lambda^m \dot \epsilon,
\end{eqnarray}
and only if $H_{\mathrm{rest}}(p, q) = 0$. Being a function of $p
$ and $q$, $H_{\mathrm{rest}}(p, q)$ will transform as a scalar, and there is no prefactor to cancel the transformation as there is for the constraints. Therefore, the action will not be invariant unless $H_{\mathrm{rest}} = 0$. This means that time evolution in general relativity consists of gauge transformations.

The last term in (\ref{Hamiltonian}) is the surface charge
\begin{equation}\label{Terashima surface charge}
J(N) = \int_{\Delta^{\infty}_{t}} \sqrt{\sigma}~(\frac{1}{8 \pi
G}N \theta -
\frac{2}{\sqrt{h}}s_{\mu}\pi^{\mu\nu}N_{\nu})~d^{d-2}x.
\end{equation}
As opposed to the procedure described above, which starts out from
the Lagrangian fomulation of general relativity, Brown and
Henneaux~\cite{Brown-Henneaux} derived the surface charge for $2+1$-dimensional space-time by demanding the Hamiltonian to have well-defined functional derivatives with respect to the canonical variables. This is
needed for the Hamiltonian to be a well-defined generator through
the Poisson bracket with the variational principle that takes boundary terms into account. Just like in the Lagrangian case, this method, called the Regge-Teitelboim method, can only determine the surface charge up to a `constant',
which they chose such that the charge vanishes for
anti-de Sitter space-time. It becomes~\cite{Brown-Henneaux}:
\begin{eqnarray}\label{surface charge again}
J(N) = \frac{1}{16 \pi G} \lim_{r \to \infty} \int \{\bar
G^{ijkl}[N^{\perp}\bar \nabla_{k}h_{ij} - \partial_{k}N^{\perp}(h_{ij} -
\bar h_{ij})] + 2 N^{i} \pi^{l}_{~i}\} dx_{l}
\end{eqnarray}
with
\begin{eqnarray}
\bar G^{ijkl} = \frac{1}{2}\sqrt{\bar h}(\bar h^{ik}\bar h^{jl} +
\bar h^{il}\bar h^{jk} - 2\bar h^{ij}\bar h^{kl})
\end{eqnarray}
where Latin indices denote parallel directions in the hypersurface-adapted frame, and the bar is used for
quantities evaluated in anti-de Sitter space. This is
equivalent~\cite{Terashima} to subtracting the anti-de Sitter
background from (\ref{Terashima surface charge}) as
\begin{equation}\label{surface charge}
J(N) = \int_{\Delta^{\infty}_{t}} ~\frac{1}{8 \pi
G}(N\sqrt{\sigma}\theta - \bar N\sqrt{\bar \sigma}\bar \theta) -
\frac{2}{\sqrt{h}}s_{\mu}\pi^{\mu\nu}N_{\nu})~d^{d-2}x,
\end{equation}
where $\bar \theta$ is the trace of the extrinsic curvature of the
$\Delta^{\infty}_{t}$ hypersurface in anti-de Sitter space. Notice that we can do this because $\bar N \sqrt{\bar \sigma}\bar \theta$ is constant in the sense that we define its derivative with respect to any of the canonical variables to vanish.

Even though the surface charges are not the usual Noether charges (see the Appendix) they do generate transformations. We have seen that the Hamiltonian in the form (\ref{Hamiltonian}) generates time translations. If we want to generate transformations in the
direction of a general vector
\begin{equation}
\xi^{\mu} = \xi^{\perp}n^{\mu} + \xi^{\parallel\mu},
\end{equation}
(which has the components $\xi^{t} = \frac{1}{N} \xi^{\perp}$,
$\xi^{a} =-\frac{N^{a}}{N}\xi^{\perp} + \xi^{\parallel a}$ in the
$(t, x^{a})$ coordinate system) all we have to do is make the
replacements $N \to \xi^{\perp}$, $N^{\mu} \to
\xi^{\parallel\mu}$. For instance, working on $h_{\mu\nu}$, we
have
\begin{eqnarray}
\{h_{\mu\nu}, H(\xi)\}
&=& \mathcal{L}_{\xi^{\perp}n}h_{\mu\nu} + h^{\alpha}_{~\mu}h^{\beta}_{~\nu}\mathcal{L}_{\xi^{\parallel}} h_{\alpha\beta} \nonumber \\
&=& \mathcal{L}_{\xi} h_{\mu\nu}.
\end{eqnarray}
As discussed at the end of Chapter 3, we can set the constraints to zero,
\begin{equation}
\mathcal{H} = 0,\ \ \ \mathcal{H}_{\mu} = 0,
\end{equation}
and replace Poisson brackets by Dirac brackets, so that the realization of the diffeomorphism algebra reduces to just the surface charges $J(\xi)$.
The group property is expressed by
\begin{equation}
\{J(\eta), J(\xi)\}^* = J([\eta, \xi]),
\end{equation}
or in case of a central extension,
\begin{equation}\label{K}
\{J(\eta), J(\xi)\}^* = J([\eta, \xi]) + K(\eta, \xi),
\end{equation}
where $K(\eta, \xi)$ is the central term. Note that $J(\xi)$ is not actually a proper generator of diffeomorphisms, since its functional derivative with respect to $h_{\mu\nu}$ and $\pi^{\mu\nu}$ is not well-defined. Therefore, we pretend that we have written the full $H(\xi)$ (including constraints) in (\ref{K}), then taken functional derivatives, and set the constraints to zero afterwards. 

The Hamiltonian with surface charge generates diffeomorphisms with either variational principle - taking into account boundary terms or ignoring them. The Hamiltonian without surface charge, on the other hand, can only be used as a generator if we neglect boundary terms. Therefore, the formulation with surface charge is the appropriate one for describing effects on the boundary.

\chapter{Geometry}

In this chapter we give a short overview of the
geometrical properties of anti-de Sitter space, which is the
maximally symmetric solution of Einstein's equations with negative
cosmological constant. We also treat the BTZ black hole \cite{BTZ}, which is a $(2+1)$-dimensional solution with negative cosmological constant and a point source at the origin. Asymptotically, the BTZ solution reduces to anti-de Sitter space, as might be expected. However, the BTZ black hole metric is actually asymptotically anti-de Sitter in a stronger sense. This will become clear further on.

\section{Anti-de Sitter Space-Time}

Anti-de Sitter space in general space-time dimension $d$ can most easily be
defined as a hyperbolic hypersurface embedded in flat
$(d+1)$-dimensional space with metric $\eta_{\mu\nu} =
\mathrm{diag}(-1, -1, 1, 1,\ldots, 1)$ via
\begin{equation}\label{ads}
\eta_{\mu\nu}X^{\mu}X^{\nu} = -l^2
\end{equation}
($\mu$, $\nu \in (1,\ldots, d+1)$). The radius of the anti-de
Sitter space is $l$, which is related to the cosmological constant
through $\Lambda = - \frac{(d-1)(d-2)}{2l^2}$. Flat space is recovered in the limit $l \to \infty$. The
definition (\ref{ads}) immediately makes clear that the isometries constitute the group $SO(d-1,2)$, which leaves invariant
$\eta_{\mu\nu}X^{\mu}X^{\nu}$. In terms of the coordinates
appearing in (\ref{ads}), the generators take the form
\begin{equation}
G_{\mu\nu} = X_{\mu}\frac{\partial}{\partial X^{\nu}} -
X_{\nu}\frac{\partial}{\partial X^{\mu}}.
\end{equation}
$G_{\mu\nu}$ is antisymmetric in its indices, so that there are
$\frac{1}{2}d(d+1)$ linearly independent generators, precisely the number of
Killing vectors of a maximally symmetric solution in $d$
dimensions.
\\
\\
\\
\\
\\
For three-dimensional anti-de Sitter space-time (AdS$_3$) we can define the
coordinates
\begin{eqnarray}
t &=& l ~\arctan\left(\frac{X^1}{X^2}\right), \nonumber \\
r &=& (X^1)^2 + (X^2)^2 -l^2, \nonumber \\
\phi &=& \arctan\left(\frac{X^3}{X^4}\right),
\end{eqnarray}
in terms of which the line element reads
\begin{equation}\label{anti-de Sitter}
ds^2 = -\left(\frac{r^2}{l^2} + 1\right) dt^2 + \left(\frac{r^2}{l^2} +
1\right)^{-1}dr^2 + r^2 d\phi^2.
\end{equation}
These coordinates have the ranges $t \in [0, 2\pi l)$, $\phi \in
[0,2\pi)$, and $r \in [0, \infty)$, i.e., both $t$ and $\phi$ are
compactified. The space-time thus contains closed timelike curves,
which is physically unacceptable because particles will be able
to influence themselves by `passing through the same point in
space-time' more than once. Therefore, we unwrap the time
coordinate $t$ to become an element of $(-\infty, \infty)$,
obtaining the so-called universal covering space of AdS$_3$. This
spacetime is commonly referred to simply as AdS$_3$. The closely
related maximally symmetric solution with positive cosmological
constant is called de Sitter space, for which the metric is
(\ref{anti-de Sitter}) with $l^2$ replaced by $-l^2$.

Locally, one can choose the coordinates 
\begin{eqnarray}
t &=& \frac{-X^2}{l^2(X^1 + X^3)} \nonumber \\
r &=& - \ln\left(\frac{l}{X^1 + X^3} \right) \nonumber \\
\phi &=& \frac{l X^4}{X^1 + X^3}
\end{eqnarray}
such that
\begin{equation}\label{Poincare}
ds^2 = -\frac{r^2}{l^2}dt^2 + \frac{l^2}{r^2}dr^2 + r^2 d\phi^2.
\end{equation}
The coordinates appearing in (\ref{Poincare}) are called
Poincar\'e coordinates. Yet another alternative is 
\begin{equation}\label{warped}
ds^2 = l^2dr^2 + e^{2r}\left(d\phi^2 - \frac{dt^2}{l^2}\right),
\end{equation}
which is obtained by making the replacement $r \to e^r$. The last two metrics show that the boundary at $r = \infty$ is a flat $(1+1)$-dimensional cylinder.
\\
\\
\\
\\
\section{The BTZ Black Hole}

The BTZ black hole is a solution of Einstein's equations with negative cosmological constant $\Lambda = -\frac{1}{l^2}$ and vanishing stress tensor everywhere except at the origin, where
there is a point source \cite{Cangemi}.
In ADM decomposition\footnote{To clean up the notation, factors of $8G$ are left implicit in $M$ and $J$.},
\begin{equation}\label{BTZ metric}
ds^2 = -N^2 dt^2 + N^{-2}dr^2 + r^2(N^{\phi}dt + d\phi)^2,
\end{equation}
where lapse $N(r)$ and angular shift $N^{\phi}(r)$ are given by
\begin{equation}
N^2(r) = -M + \frac{r^2}{l^2} + \frac{J^2}{4r^2},\ \ \ N^{\phi}(r)
= -\frac{J}{2r^2}.
\end{equation}
It can be verified that at large $r$, the BTZ black hole solution approaches three-dimensional anti-de Sitter space. This was to be expected, since as $r \to \infty$, we are increasingly far removed from the delta function source. However,  the BTZ black hole actually obeys more restrictive asymptotic conditions which will be discussed in the next chapter.

Unlike AdS$_3$, the BTZ black hole metric has only two Killing vectors, $\frac{d}{d t}$ and $\frac{d}{d \phi}$. That these are Killing vectors can immediately be seen from the metric
(\ref{BTZ metric}), which depends neither on $t$ nor on $\phi$. The
conserved quantities associated with time translation and
rotational symmetry are the mass $M$ and angular momentum $J$ of
the black hole, respectively. The fact that $\frac{d}{dt}$ is a symmetry of (\ref{BTZ metric}) means that the BTZ black hole is a stationary solution.

The inner and outer horizons, where spacelike and timelike vectors interchange roles, are located at the roots of the lapse function:
\begin{equation}
r_{\pm} = l\left(\frac{M}{2}\left(1 \pm \sqrt{1 -
\left(\frac{J}{lM}\right)^2}\right)\right)^{\frac{1}{2}}.
\end{equation}
The BTZ black hole is a spinning black hole.
Spinning black holes have a region outside the outer horizon in which particles cannot remain at rest, called the ergosphere. The outer limit of the BTZ ergosphere is located at the surface of infinite redshift $r = l\sqrt{M}$, where $g_{tt}$ vanishes.
The extreme black hole is obtained when inner and outer horizons
coincide, i.e.,
\begin{equation}
|J| = lM.
\end{equation}
We also see that $M$ is defined such that it vanishes as $r_{\pm} \to
0$, so that the zero point of energy is reached when the black
hole disappears. In the opposite extreme, for $M \ne 0$, $l \to \infty$, we
are left with only the inside. The BTZ black hole carries no
charge; the charged $2+1$ black hole is rather different in that
its curvature is not constant\ \cite{BTZH}. Substituting $J = 0$, $M
= -1$ in the BTZ metric gives back AdS$_3$. However, AdS$_3$
cannot be obtained from the BTZ black hole vacuum \mbox{($M = 0, J = 0$)}
in a continuous way, since the negative mass solutions with $M$
between $0$ and $-1$ possess a naked singularity, which is
physically unacceptable\ \cite{BTZH}. This means that the BTZ black hole cannot decay physically into AdS$_3$. 
\\
\\

Since the BTZ black hole is also a solution of the
Einstein equations with vanishing stress-energy tensor (except at
the origin), it can however only differ from AdS$_3$ by some
discrete global identifications. This is because in three space-time dimensions, all local information is stored in the Einstein tensor $G_{\mu\nu} = R_{\mu\nu} - \frac{1}{2}Rg_{\mu\nu}$, and the latter coincides for AdS$_3$ and the BTZ black hole outside the source. $R_{\mu\nu\rho\sigma}$ contains no additional information due to the relation
\begin{equation}
R_{\mu\nu\rho\sigma} = \epsilon_{\mu\nu\kappa}\epsilon_{\rho\sigma\lambda}G^{\lambda\kappa}
\end{equation}
which holds in three dimensions. Indeed, it can be verified that $R_{\mu\nu}$ and $R_{\mu\nu\sigma\rho}$ each have six independent components in $d = 3$.
In\ \cite{BTZH}, a procedure is
described to obtain the BTZ black hole through identifications in
the universal covering space of AdS$_3$. The identification group
turns out to be a discrete subgroup of $SO(2,2)$, which is very
natural, because in this way, the manifold with the
identifications inherits a well-defined local structure.

\chapter{Central Charge of Asymptotically AdS$_3$ Space-Times}

The presence of a classical central charge in the asymptotic
symmetry algebra of certain three-dimensional asymptotically
anti-de Sitter space-times was first noted by Brown and Henneaux
in $1986$\ \cite{Brown-Henneaux}. Often, central charges only
arise upon quantization, for instance due to normal ordering of
lowering and raising operators, as explained in section 2.3. The fact that a central charge arises already classically in this case has to do with the fact that the asymptotic Killing vectors do not all preserve the boundary metric exactly. 
The asymptotic symmetries that are actually exact symmetries of the AdS$_3$ background do not lead to a central extension. 
It is necessarily the case that the boundary metric is not preserved by all asymptotic symmetries, because the isometry group turns out to be extended to an infinite-dimensional group `at the boundary' - the conformal group in $1+1$ dimensions. We have seen before that the conformal group in higher dimensions has only a finite number of generators, making this a phenomenon specific to a three-dimensional bulk theory.

Asymptotic symmetries may be naively defined as any diffeomorphisms leading to corrections that are term by term of lower order in $r$ than the original metric. If we start from the metric (\ref{Poincare}), these include all diffeomorphisms that result (locally) in 
\begin{equation}\label{rescale}
ds^2 = - \left(\frac{r^2}{l^2} + O(r)\right) dt^2 + \left(\frac{l^2}{r^2} + O\left(\frac{1}{r^3}\right) \right) dr^2 + (r^2 + O(r)) d\phi^2. 
\end{equation}
Brown and Henneaux translated (\ref{rescale}) into conditions on the asymptotic metric by recognizing that at least the AdS$_3$ isometry group $SO(2,2)$ should be among the asymptotic symmetries. $SO(2,2)$ transformations are clearly asymptotic symmetries of AdS$_3$ in the above sense. Then, if we first perform an asymptotic symmetry transformation on AdS$_3$ that is not contained in $SO(2,2)$, obtaining a different metric that takes the form (\ref{rescale}), this metric should still be asymptotically invariant under $SO(2,2)$ if we want the transformations to form a group. In the dynamical description of these transformations, another condition on the asymptotic metric is that its asymptotic symmetries should have finite canonical generators $J(\xi)$.

The asymptotic metric used by Brown and Henneaux takes the form:
\begin{eqnarray}\label{Brown-Henneaux metric}
g_{tt} &=& - \frac{r^2}{l^2} + O(1) \nonumber \\
g_{tr} &=& O(\frac{1}{r^3}) \nonumber \\
g_{t\phi} &=& O(1) \nonumber \\
g_{rr} &=& \frac{l^2}{r^2} + O(\frac{1}{r^4}) \nonumber \\
g_{r\phi} &=& O(\frac{1}{r^3}) \nonumber \\
g_{\phi\phi} &=& r^2 + O(1)
\end{eqnarray}
where $t \in (-\infty, \infty)$, $r \in [0, \infty)$, and $\phi
\in [0, 2\pi)$. From now on, `asymptotically anti-de Sitter' will be understood to mean (\ref{Brown-Henneaux metric}).

This chapter is organized as follows. In section 6.1 we begin by
deriving the diffeomorphisms preserving 
(\ref{Brown-Henneaux metric}). These will include
$SO(2,2)$ as a subgroup, as the asymptotic
metric was selected to be invariant under this group. Using the $SO(2,2)$ subgroup of the asymptotic Killing vectors, we come back to the form of the asymptotic metric in section 6.2. We show that demanding invariance under $SO(2,2)$ leads to conditions that are not quite as strict as (\ref{Brown-Henneaux metric}). In section 6.3, the additional restrictions are shown to follow from the demand that the surface charges be finite. The main part of
our discussion comes in section 6.4, where we calculate the
central charge associated with the asymptotic symmetry group of
(\ref{Brown-Henneaux metric}) in the canonical formalism. Section 6.5 contains an alternative derivation of the central charge from the work of Balasubramanian and Kraus \cite{BK}. 

\section{Asymptotic Killing Vectors}

In this section we derive the asymptotic isometries of (\ref{Brown-Henneaux metric}). These have a natural $SO(2,2)$ subgroup, as the asymptotic metric was obtained under the very condition that they be $SO(2,2)$-invariant.

Without loss of generality, we can write the candidate vectors in
a Laurent expansion in $r$. Denoting equal or higher powers of $r$
by $o$ and equal or lower powers by $O$\footnote{We do not write
out more terms due to some prior knowledge on what the Killing
vectors will turn out to look like from \cite{Brown-Henneaux} and
\cite{Strominger}.}:
\begin{eqnarray}\label{Laurent expand}
\xi^t &=& o(r) + A(t, \phi) + \frac{B(t, \phi)}{r} + \frac{C(t, \phi)}{r^2} + \frac{D(t, \phi)}{r^3} + O(\frac{1}{r^4}), \nonumber \\
\xi^r &=& o(r^2) + E(t, \phi)r + F(t, \phi) + O(\frac{1}{r}), \nonumber \\
\xi^{\phi} &=& o(r) + G(t, \phi) + \frac{H(t, \phi)}{r} + \frac{I(t, \phi)}{r^2} + \frac{J(t, \phi)}{r^3} + O(\frac{1}{r^4}).
\end{eqnarray}
Applying Lie transport with these vectors $\xi^{\mu}$ to
$g_{\mu\nu}$ and demanding that (\ref{Brown-Henneaux metric}) is preserved, results in
\begin{eqnarray}\label{conditions}
B = D = F = H = J = 0, \nonumber \\
2 C = - l^4 \partial_t E, \nonumber \\
\partial_{\phi}A = l^2 \partial_t G, \nonumber \\
2 I = l^2 \partial_{\phi}E, \nonumber \\
E = - \partial_{\phi}G = - \partial_t A.
\end{eqnarray}
and all higher order terms should vanish.

The fact that $[l^2\partial_t^2 - \partial_{\phi}^2]A(t,\phi) = 0$ allows us to write $A(t,\phi)$ as $l(\eta^{+} + \eta^{-})$, where
$\partial_{\pm}\eta^{\mp} = 0$ and $\partial_{\pm} \equiv
\frac{1}{2}(l \partial_{t} \pm \partial_{\phi})$. The other components then follow uniquely from the relations (\ref{conditions}), and it can be verified that the asymptotic Killing vectors become
\begin{eqnarray}\label{vectors}
\xi^{t} &=& l(\eta^{+} + \eta^{-}) + \frac{l^3}{2 r^2}(\partial_{+}^{2}\eta^{+} +
\partial_{-}^{2}\eta^{-}) + O(\frac{1}{r^4}), \nonumber \\
\xi^{r} &=& -r(\partial_{+}\eta^{+} + \partial_{-}\eta^{-}) + O(\frac{1}{r}), \nonumber \\
\xi^{\phi} &=& \eta^{+} - \eta^{-} - \frac{l^2}{2 r^2}(\partial_{+}^{2}\eta^{+} -
\partial_{-}^{2}\eta^{-}) + O(\frac{1}{r^4}).
\end{eqnarray}
We can see that the $\eta^+$ and $\eta^-$ transformations do not talk to each other,
\begin{eqnarray}
\big[ \xi(\eta^-_1), \xi(\eta^-_2) \big] = \xi(\eta^-_1\partial_-\eta^-_2 - \eta^-_2\partial_-\eta^-_1) \nonumber \\
\big[ \xi(\eta^+_1), \xi(\eta^+_2) \big] = \xi(\eta^+_1\partial_+\eta^+_2 - \eta^+_2\partial_+\eta^+_1). 
\end{eqnarray}
Moreover, it is clear that we are dealing with an infinite-dimensional group.
Following Strominger \cite{Strominger} we can then let $l_{n}$ and $\bar
l_{n}$ denote the generators of the diffeomorphisms with $\eta^{+}
= e^{in(\frac{t}{l} + \phi)}$, $\eta^{-} = 0$ and $\eta^{-} =
e^{in(\frac{t}{l} - \phi)}$, $\eta^{+} = 0$, respectively. The resulting Lie bracket algebra is the loop algebra
\begin{eqnarray}
\left[ l_m, l_n \right] &=& 2i(m-n)l_{m+n} \nonumber \\
\left[ \bar l_m, \bar l_n \right] &=& 2i(m-n)\bar l_{m+n} \nonumber \\
\left[ l_m, \bar l_n \right] &=& 0,
\end{eqnarray}
and we have indeed found the infinite-dimensional conformal group on the \mbox{$(1+1)$-dimensional} boundary.

\newpage

The $so(2,2)$ subalgebra is spanned by $l_{-1}, l_{0}, l_{1}, \bar
l_{-1}, \bar l_{0}, \bar l_{1}$. These take the form
\begin{eqnarray}
l_{-1}^t &=& l ~e^{-i(\frac{t}{l} + \phi)} - \frac{l^3}{2r^2}~e^{-i(\frac{t}{l} + \phi)} \nonumber \\
l_{-1}^r &=& ir ~e^{-i(\frac{t}{l} + \phi)} \nonumber \\
l_{-1}^{\phi} &=& e^{-i(\frac{t}{l} + \phi)} + \frac{l^2}{2r^2}~e^{-i(\frac{t}{l} + \phi)} \nonumber \\
\\
l_{0}^t &=& l \nonumber \\
l_{0}^r &=& 0 \nonumber \\
l_{0}^{\phi} &=& 1 \nonumber \\
\\
l_{1}^t &=& l ~e^{i(\frac{t}{l} + \phi)} - \frac{l^3}{2r^2}~e^{i(\frac{t}{l} + \phi)} \nonumber \\
l_{1}^r &=& -ir~e^{i(\frac{t}{l} + \phi)} \nonumber \\
l_{1}^{\phi} &=& e^{i(\frac{t}{l} + \phi)} + \frac{l^2}{2r^2}~e^{i(\frac{t}{l} + \phi)} \nonumber \\
\\
\bar l_{-1}^t &=& l ~e^{-i(\frac{t}{l} - \phi)} - \frac{l^3}{2r^2}~e^{-i(\frac{t}{l} - \phi)} \nonumber \\
\bar l_{-1}^r &=& ir ~e^{-i(\frac{t}{l} - \phi)} \nonumber \\
\bar l_{-1}^{\phi} &=& -e^{-i(\frac{t}{l} - \phi)} - \frac{l^2}{2r^2}~e^{-i(\frac{t}{l} - \phi)} \nonumber \\
\\
\bar l_{0}^t &=& l \nonumber \\
\bar l_{0}^r &=& 0 \nonumber \\
\bar l_{0}^{\phi} &=& -1 \nonumber \\
\\
\bar l_{1}^t &=& l ~e^{i(\frac{t}{l} - \phi)} - \frac{l^3}{2r^2}~e^{i(\frac{t}{l} - \phi)} \nonumber \\
\bar l_{1}^r &=& -ir~e^{i(\frac{t}{l} - \phi)} \nonumber \\
\bar l_{1}^{\phi} &=& -e^{i(\frac{t}{l} - \phi)} - \frac{l^2}{2r^2}~e^{i(\frac{t}{l} - \phi)}
\end{eqnarray}

\newpage

\section{The Asymptotic Metric}

We can now reverse the process, and use the $SO(2,2)$ vectors given above to derive asymptotic conditions for the metric. These conditions will not yet be quite as restrictive as (\ref{Brown-Henneaux metric}). We thus find an asymptotic metric that is consistent under $SO(2,2)$, and allows for an even larger asymptotic symmetry group \mbox{than (\ref{Brown-Henneaux metric})}.

To derive the form the asymptotic metric should take, we start by
looking at the transformations generated by $l_{0}$ and $\bar l_{0}$:
\begin{eqnarray}
\delta_{l_0} g_{\mu\nu} &=& (l \partial_t + \partial_{\phi}) g_{\mu\nu} \nonumber \\
\delta_{\bar l_0} g_{\mu\nu} &=& (l \partial_t - \partial_{\phi}) g_{\mu\nu}.
\end{eqnarray}
As these variations are of the same order of $r$ as the metric
itself, this tells us that any exact leading order terms in $r$ should
not depend on $t$ or $\phi$.
Lie transport of $g_{tt}$ with $l_{\pm 1}$ gives
\begin{eqnarray}
\delta_{l_{\pm 1}} g_{tt}&=&2 g_{tt} \partial_{t}l_{\pm 1}^{t} + 2 g_{rt}\partial_{t}l_{\pm 1}^{r} + 2 g_{t\phi}\partial_{t}l_{\pm 1}^{\phi} + l_{\pm 1}^{t}\partial_{t}g_{tt} + l_{\pm 1}^{r}\partial_{r}g_{tt} + l_{\pm 1}^{\phi}\partial_{\phi}g_{tt} \nonumber \\
&=& [2 (\pm i \mp \frac{il^2}{2r^2})g_{tt} + 2 \frac{r}{l} g_{tr}+ 2(\pm \frac{i}{l} \pm \frac{il}{2r^2})g_{t\phi} + (l - \frac{l^3}{2r^2})\partial_{t}g_{tt} \nonumber \\
&& \ \ \ \ \ \ \ \ \ \ \ \ \ \ \  \ \mp ir\partial_{r}g_{tt} + (1 + \frac{l^2}{2r^2})\partial_{\phi}g_{tt}] ~e^{\pm i(\frac{t}{l} + \phi)}.
\end{eqnarray}
We can add up the transformations under $l_{1}$ and $l_{-1}$ for $\frac{t}{l} + \phi = 2\pi n , n \in Z$:
\begin{eqnarray}
\delta_{(l_{1} + l_{-1})}g_{tt} = 4 \frac{r}{l} g_{tr} + (2l - \frac{l^3}{r^2})\partial_{t}g_{tt} + (2 + \frac{l^2}{r^2})\partial_{\phi}g_{tt}.
\end{eqnarray}
This tells us that if the leading order of $g_{tt}$ is to be
exactly preserved (so that, as we have just seen, it does not
depend on $t$ or $\phi$) $g_{tr}$ should be of at least two orders
lower than $g_{tt}$. Subtracting one from the other, we get
\begin{eqnarray}
\delta_{(l_{1} - l_{-1})}g_{tt} = (4i - \frac{2l^2}{r^2})g_{tt} + (\frac{4i}{l} + \frac{2il}{r^2})g_{t\phi} - 2ir\partial_{r}g_{tt}
\end{eqnarray}
showing that, if $g_{t\phi}$ is lower order, $g_{tt}$ obeys
\begin{equation}
r\partial_{r}g_{tt} = 2 g_{tt}
\end{equation}
to leading order in $r$.
Similarly (still with $\frac{t}{l} + \phi = 2\pi n$),
\begin{equation}
\delta_{(l_{1} - l_{-1})}g_{t\phi} = (2il - \frac{il^3}{r^2})g_{tt} + 4ig_{t\phi} + (\frac{2i}{l} + \frac{il}{r^2})g_{\phi\phi} - 2ir\partial_{r}g_{t\phi}
\end{equation}
shows that to leading order
\begin{equation}\label{t and phi}
g_{\phi\phi} = - l^2 g_{tt}
\end{equation}
if $g_{t\phi}$ is of lower order than $g_{tt}$ and $g_{\phi\phi}$,
and the latter are of the same order. In fact, suppose that either
$g_{tt}$ or $g_{\phi\phi}$ were of highest order, then that
component of the metric should vanish completely not to give high
order corrections to $g_{t\phi}$. Therefore, as long as
$g_{t\phi}$ is of lower order than both, $g_{tt}$ and
$g_{\phi\phi}$ should indeed have the same order, and (\ref{t and
phi}) should hold. The transformation
\begin{equation}
\delta_{(l_{1} - l_{-1})}g_{rr} = -4ig_{rr} - 2ir \partial_{r}g_{rr}
\end{equation}
gives
\begin{equation}
r\partial_{r}g_{rr} = -2g_{rr}
\end{equation}
for the exact leading order part of $g_{rr}$. As a final example, we look at the
transformation of $g_{\phi\phi}$ under Lie transport with $l_{1} - l_{-1}$:
\begin{equation}
\delta_{(l_{1} - l_{-1})}g_{\phi\phi} = (4il - \frac{2il^3}{r^2})g_{t\phi} + (4i + \frac{2il^2}{r^2})g_{\phi\phi} - 2ir\partial_{r}g_{\phi\phi}.
\end{equation}
From this we derive
\begin{equation}
r\partial_{r}g_{\phi\phi} = 2g_{\phi\phi}
\end{equation}
for any exact leading order of $g_{\phi\phi}$, as long as $g_{t\phi}$ is
lower order. Proceeding in the same way for different combinations
of $l_{\pm 1}$ and $\bar l_{\pm 1}$ and $\frac{t}{l} = (m+n)\pi$,
$\phi = (n-m)\pi$, $n, m \in Z$, so that all exponentials become
equal to $1$, we can derive many more relations, each of which is obeyed by the asymptotic metric
\begin{eqnarray}\label{ansatz}
g_{tt} &=& - \frac{r^2}{l^2} + O(1) \nonumber \\
g_{tr} &=& O\left(\frac{1}{r}\right) \nonumber \\
g_{t\phi} &=& O(1) \nonumber \\
g_{rr} &=& \frac{l^2}{r^2} + O\left(\frac{1}{r^3}\right) \nonumber \\
g_{r\phi} &=& O\left(\frac{1}{r}\right) \nonumber \\
g_{\phi\phi} &=& r^2 + O(1).
\end{eqnarray}
These fall-off conditions are consistent under $SO(2,2)$, and the vectors preserving (\ref{ansatz}) obey
\begin{eqnarray}\label{relations}
B = F = H = 0 \nonumber \\
\partial_{\phi}A = l^2 \partial_t G, \nonumber \\
E = - \partial_{\phi}G = - \partial_t A.
\end{eqnarray}
There are no further restrictions on $C, D, I$, and $J$. That the
anti-de Sitter group is part of the asymptotic symmetries is
reflected by the fact that all of these relations also hold for
the vectors (\ref{vectors}). The vectors obeying (\ref{relations}) \mbox{look like}
\begin{eqnarray}
\xi^{t} &=& l(\eta^{+} + \eta^- )+ O(\frac{1}{r^2}), \nonumber \\
\xi^{r} &=& -r(\partial_{+}\eta^{+} + \partial_{-}\eta^{-}) + O(\frac{1}{r}), \nonumber \\
\xi^{\phi} &=& \eta^{+} - \eta^{-} + O(\frac{1}{r^2}).
\end{eqnarray}
Thus, it seems that we have found an even larger asymptotic symmetry group than the conformal group. However, we will see in the next section that if we introduce Hamiltonian dynamics, the boundary conditions (\ref{ansatz}) need to be sharpened.

We have seen that the leading order terms of a metric that has
$SO(2,2)$ in its asymptotic symmetry group are those of anti-de
Sitter space-time. This is by no means surprising. However, note
that, vice versa, the fact that $SO(2,2)$ is part of the
asymptotic symmetries does \emph{not} follow from the leading
orders of the metric being those of anti-de Sitter space. We can
either weaken or strengthen the boundary conditions in such a way
that they are no longer preserved by all $SO(2,2)$ vectors. As an
example, consider
\begin{eqnarray}
g_{tt} &=& -\frac{r^2}{l^2} + O(r) \nonumber \\
g_{tr} &=& O(r) \nonumber \\
g_{t\phi} &=& O(r) \nonumber \\
g_{rr} &=& \frac{l^2}{r^2} + O\left(\frac{1}{r^3}\right) \nonumber \\
g_{r\phi} &=& O(r) \nonumber \\
g_{\phi\phi} &=& r^2 + O(r).
\end{eqnarray}
It can be verified that the vectors (\ref{Laurent expand}) should
have $\partial_t E = \partial_{\phi}E = 0$ for these asymptotic
conditions to be consistent. This is the case for $l_0$ and $\bar
l_0$ (which are symmetries whenever the leading order components of the
metric do not depend on $t$ or $\phi$), but not for $SO(2,2)$
vectors in general.

\section{Finiteness of the Surface Charges}

So far, we have treated the asymptotic symmetries outside a dynamical context. If we introduce Hamiltonian mechanics in the general relativistic setting, the asymptotic symmetries are generated by the surface charges $J(\xi)$. We show that the demand that these generators be well-defined leads to additional conditions on the asymptotic metric. 

For the asymptotics (\ref{ansatz}), the
surface charge (\ref{surface charge again}) given in Chapter 4
simplifies to
\begin{eqnarray}
J(\xi) = \frac{1}{16\pi G} \lim_{r \to \infty} \int \{\frac{l}{r}\xi^{\perp} + \frac{r^3}{l^3}\xi^{\perp}(g_{rr} - \frac{l^2}{r^2}) + \frac{1}{l}(\frac{1}{r}\xi^{\perp} + \partial_{r}\xi^{\perp})(g_{\phi\phi} - r^2) 
\nonumber \\
+ \frac{1}{l}(\xi^{\perp}\partial_{\phi}g_{r\phi} - g_{r\phi}\partial_{\phi}\xi^{\perp}) + 2 \xi^{\parallel \phi}\pi^{~r}_{\phi} \} d\phi.
\end{eqnarray}
The term $\frac{1}{l}(\xi^{\perp}\partial_{\phi}g_{r\phi} - g_{r\phi}\partial_{\phi}\xi^{\perp})$ goes to zero for (\ref{Brown-Henneaux metric}), because for that metric it is of $O\left(\frac{1}{r^2}\right)$. Here, it is of $O(1)$, which would not pose a difficulty if it were not for the fact that $g_{r\phi}$ obtained via Lie transport of the anti-de Sitter metric with an allowed vector $\eta^{\mu}$ (hence yielding an allowed metric),
\begin{eqnarray}
g_{r\phi} &=& \bar g_{r\phi} + \mathcal{L}_{\eta}\bar g_{r\phi} \nonumber \\
&=& \bar g_{rr} \partial_{\phi} \eta^r + \bar g_{\phi\phi} \partial_r \eta^{\phi},
\end{eqnarray}
contains unspecified orders of $O\left(\frac{1}{r}\right)$ due to the arbitrary $O\left(\frac{1}{r^2}\right)$ term \mbox{in $\eta^{\phi}$}. 

\noindent The same problem arises with $\frac{1}{l}(\frac{1}{r}\xi^{\perp} + \partial_{r}\xi^{\perp})(g_{\phi\phi} - r^2)$, for
\begin{eqnarray}
g_{\phi\phi} &=& \bar g_{\phi\phi} + \mathcal{L}_{\eta}\bar g_{\phi\phi} \nonumber \\
&=& r^2 + 2\bar g_{\phi\phi} \partial_{\phi}\eta^{\phi} + \eta^r \partial_r \bar g_{\phi\phi}
\end{eqnarray}
has an undetermined $O(1)$ term, while $\frac{1}{r}\xi^{\perp} +
\partial_{r}\xi^{\perp}$ is of the same order, so that the product
does not vanish in the limit $r \to \infty$. Similarly, the term
$2 \xi^{\phi} \pi ^{~r}_{\phi}$ contains unspecified orders of
$O(1)$ coming from $O\left( \frac{1}{r^2} \right)$ in $\eta^t$, since $\pi^{~r}_{\phi}$ is equal to leading order in $r$ to $g_{t\phi}$ \cite{Brown-Henneaux}. It is clear that we need more restrictions on the asymptotic Killing vectors in order to have well-defined surface charges. These
restrictions will come from stricter fall-off conditions for the
metric. Specifically, $2 C = - l^4 \partial_t E$ comes from
$g_{tr}$ being of order $O\left(\frac{1}{r^2}\right)$ rather than
$O\left(\frac{1}{r}\right)$, and $2 I = l^2 \partial_{\phi}E$
comes from $g_{r\phi}$ being of order
$O\left(\frac{1}{r^2}\right)$. Moreover, the term
$\frac{r^3}{l^3}\xi^{\perp}(g_{rr} - \frac{l^2}{r^2})$ diverges
unless $g_{rr} = \frac{l^2}{r^2} + O\left(\frac{1}{r^4}\right)$,
after which consistency even requires $g_{tr} =
O\left(\frac{1}{r^3}\right)$ and $g_{r\phi} =
O\left(\frac{1}{r^3}\right)$. This has as a consequence that $D =
J = 0$ in (\ref{Laurent expand}) and finally leaves us with
(\ref{Brown-Henneaux metric}).

\section{Boundary Dynamics and the Central Charge}

We mentioned that the Poisson bracket algebra of the asymptotic symmetries of (\ref{Brown-Henneaux metric}) turns out to be centrally extended. The resulting algebra looks like
\begin{eqnarray}\label{Virasoro algebra}
\{J(l_{m}), J(l_{n})\} &=& 2i(m - n)J(l_{m + n}) + \frac{ic}{3}(m^3 - m)\delta_{m +
n,0}, \nonumber \\
\{J(\bar l_{m}),J(\bar l_{n})\} &=& 2i(m - n)J(\bar l_{m + n}) +
\frac{i\bar c}{3}(m^3 - m)\delta_{m + n,0}, \nonumber \\
\{ J(l_{m}),J(\bar l_{n})\} &=& 0,
\end{eqnarray}
corresponding to two copies of the Virasoro algebra with an as yet unknown central
charge $c$. In the present section we calculate this central charge by making use of the canonical formalism introduced in Chapter 4. We will also see that
$c = \bar c$.

As discussed in the previous section, the surface charge $J(\xi)$ with anti-de Sitter background subtracted takes the form
\begin{equation}\label{simplified surface charge}
J(\xi) = \frac{1}{16\pi G} \lim_{r \to \infty} \int d\phi \{\frac{l}{r}\xi^{\perp} + \frac{r^3}{l^3}\xi^{\perp}(g_{rr} - \frac{l^2}{r^2}) + \frac{1}{l}(\frac{1}{r}\xi^{\perp} + \partial_{r}\xi^{\perp})(g_{\phi\phi} - r^2) + 2 \xi^{\parallel \phi}\pi^{~r}_{\phi} \},
\end{equation}
for the given metric (\ref{Brown-Henneaux metric}). Since the integration is performed over the $r \rightarrow \infty$
boundary, $J(\xi)$ vanishes for any asymptotic Killing vector that is of
$O(\frac{1}{r^4})$ in its $t$ and $\phi$ components, and of
$O(\frac{1}{r})$ in its $r$ component, so that the actual group of
asymptotic symmetries may be defined as the factor group obtained by
identifying vectors that differ only in these orders in $\frac{1}{r}$ \cite{Brown-Henneaux}.

As discussed at the end of Chapter 4, the central term is given
by the difference between the Dirac bracket algebra of the surface
charges and the charge evaluated on the Lie bracket of two
asymptotic symmetries:
\begin{equation}\label{Dirac bracket algebra}
\{J(\xi), J(\eta)\}^{\ast} = J([\xi, \eta]) + K(\xi, \eta).
\end{equation}
$K(\xi, \eta)$ is the central term.
Now, the value of the charge on the surface deformed by $\eta^{\mu}$ is:
\begin{eqnarray}\label{charge on deformed surface}
J(\xi) + \delta_{\eta}J(\xi)
&=& J(\xi) + \{J(\xi), J(\eta)\}^{\ast} \nonumber \\
&=& J(\xi) + J([\xi, \eta]) + K(\xi, \eta).
\end{eqnarray}
If we start out with the AdS$_3$ metric $\bar g_{\mu\nu}$,
then both $J(\xi) = 0$ and $J([\xi, \eta]) = 0$ (substituting $g_{\mu\nu} =
\bar g_{\mu\nu}$ in the expression\ (\ref{simplified surface charge}) yields zero
for any vector), and we are left with\footnote{The Dirac bracket does not vanish, because $g_{\mu\nu}$ is treated as a variable inside the bracket.}
\begin{equation}\label{central charge is surface charge}
K(\xi, \eta) = \delta_{\eta}J(\xi).
\end{equation}
This can be evaluated by substituting
\begin{equation}
g_{\mu\nu} = \bar g_{\mu\nu} + \mathcal{L}_{\eta}\bar g_{\mu\nu}
\end{equation}
in the expression for $J(\xi)$. Since the surface charge vanishes for anti-de Sitter space, we see that $K(\xi, \eta)$ would diverge if we had not demanded the surface charges to be finite for all asymptotic Killing vectors. The $J(\xi)$ would then no longer have formed a projective representation of the asymptotic symmetries. 

If we choose $\xi = l_m$ and $\eta = l_n$ in (\ref{central charge is surface charge}), we have, to leading
order in $r$:
\begin{eqnarray}
\xi^{\perp} &=& N l_{m}^t \simeq \frac{r}{l}\left(l - \frac{m^2 l^3}{2r^2}\right)e^{im(\frac{t}{l} + \phi)},  \\
\xi^{\parallel\phi} &=& l_m^{\phi} + N^{\phi} l_m^t \simeq (1 + \frac{m^2 l^2}{2r^2})e^{im(\frac{t}{l} + \phi)}, \\
g_{rr} &=& (\frac{r^2}{l^2} + 1)^{-1} + 2\bar g_{rr} \partial_r l_{n}^r + L_n^r \partial_r \bar g_{rr} \nonumber \\
&=& (\frac{r^2}{l^2} + 1)^{-1} + 2(\frac{r^2}{l^2} + 1)^{-1} \partial_r (-inr e^{in(\frac{t}{l} + \phi)}) + -inr e^{in(\frac{t}{l} + \phi)}\partial_r (\frac{r^2}{l^2} + 1)^{-1} \nonumber \\
&\simeq& \frac{l^2}{r^2} - \frac{l^4}{r^4} - 2in\frac{l^4}{r^4}e^{in(\frac{t}{l} + \phi)},  \\
g_{\phi\phi} &=& r^2 + 2\bar g_{\phi \phi} \partial_{\phi}L_n^{\phi} + l_n^r \partial_r \bar g_{\phi\phi} \nonumber \\
&=& r^2 + 2r^2 \partial_{\phi}(1 + \frac{n^2 l^2}{2r^2})e^{in(\frac{t}{l} + \phi)} - 2inr^2 e^{in(\frac{t}{l} + \phi)} \nonumber \\
&=& r^2 + il^2n^3e^{in(\frac{t}{l} + \phi)},
\end{eqnarray}
and, also to the relevant orders in $r$, $\pi^{~r}_{\phi} = g_{t\phi}$ ~\cite{Brown-Henneaux}, with
\begin{eqnarray}
g_{t\phi} &=& \bar g_{tt} \partial_{\phi} l_n^t + \bar g_{\phi\phi}\partial_t l_n^{\phi} \nonumber \\
&=& -(\frac{r^2}{l^2} + 1)\partial_{\phi}(l - \frac{n^2 l^3}{2r^2})e^{in(\frac{t}{l} + \phi)} + r^2 \partial_t (1 + \frac{n^2 l^2}{2r^2})e^{in(\frac{t}{l} + \phi)} \nonumber \\
&=& il(n^3 - n + \frac{n^3 l^2}{2r^2})e^{in(\frac{t}{l} + \phi)}.
\end{eqnarray}
For $\bar l_m$, $\bar l_n$ we obtain similar results. It can be
verified that $J(\xi)$ integrates to zero unless $m = -n$, and
that, accordingly, the central term is\footnote{Brown and Henneaux
used the aforementioned relation $[l^2\partial_t^2 - \partial_{\phi}^2]A(t,
\phi) = 0$ \mbox{(cf. (\ref{conditions}))}  to Fourier decompose the asymptotic Killing vectors,
yielding four sets of generators $A_n, B_n, C_n, D_n$, with
$A(t,\phi)$ equal to $l\cos\frac{nt}{l}\cos n\phi$,
$l\sin\frac{nt}{l}\sin n\phi$, $l\sin\frac{nt}{l}\cos n\phi$, and
$l\cos\frac{nt}{l}\sin n\phi$, respectively. The change of basis
from $A_n$, $B_n$, $C_n$ and $D_n$ to the $l_n$, $\bar l_n$ used
by Strominger, is as follows:
\begin{eqnarray}
l_n &=& A_n - B_n + iC_n +iD_n, \nonumber \\
\bar l_n &=& A_n + B_n +iC_n - iD_n \nonumber
\end{eqnarray}
(\cite{Brown-Henneaux} contains a sign error for $B_n^r$).
In Brown-Henneaux's notation, the only nonvanishing central terms are
\begin{eqnarray}
K(A_n, C_m) &=& 2\pi lm(m^2 - 1)\delta_{|n|, |m|}, \nonumber \\
K(B_n, D_m) &=& -2\pi lm(m^2 - 1)\delta_{|n|, |m|}. \nonumber
\end{eqnarray}
}
\begin{equation}\label{central term in canonical algebra}
K(l_{m}, l_{n}) =  K(\bar l_{m}, \bar l_{n}) = \frac{i l}{2G} (m^3 - m) \delta_{m+n, 0}.
\end{equation}
The central charge appearing in (\ref{Virasoro algebra}) thus becomes
\begin{equation}\label{central}
c = \bar c = \frac{3 l}{2 G}.
\end{equation}
We have found a classical central charge in the Poisson bracket algebra of the asymptotic symmetry group of three-dimensional asymptotically anti-de Sitter space-times. Since the central charge only shows up after introducing Poisson brackets, the central element does not generate any transformation.

The above derivation has immediately shown that the asymptotic
symmetries for which $\mathcal{L}_{\eta}\bar g_{\mu\nu} = 0$ have no associated central term. Hence, the $SO(2,2)$ subalgebra of
diffeomorphisms generating isometries of anti-de Sitter space
is not centrally extended.

Since we obtained two copies of the Virasoro
algebra, it seems possible that the boundary dynamics we have
described can be captured in a $(1+1)$-dimensional conformal field theory. Indeed, it
was shown in \cite{Coussaert} that the boundary degrees of freedom
correspond (up to zero modes) to those of Liouville theory.
Analogously, the boundary fluctuations in AdS$_3$ supergravity have
a dual interpretation within super-Liouville theory \cite{AdS
supergravity, Bautier, Maoz}. We summarize the derivation of \cite{Coussaert} in chapter 7.

\section{Brown-York Stress Tensor}

Since the original calculation by Brown and Henneaux, there have been many alternative derivations of the central charge of asymptotically AdS$_3$. We discuss one approach by Balasubramanian and Kraus\ \cite{BK}, who looked at the
transformation properties of the Brown-York stress-energy tensor\
\cite{Brown-York}:
\begin{equation}\label{Brown-York tensor}
T_{\mu\nu} = \frac{2}{\sqrt{-\gamma}} \frac{\delta S}{\delta \gamma^{\mu\nu}}.
\end{equation}
The tensor is called `quasilocal' because it involves the induced
boundary metric $\gamma_{\mu\nu} = g_{\mu\nu} - n_{\mu}n_{\nu}$
rather than $g_{\mu\nu}$ itself. Since the Brown-York tensor
normally diverges as the boundary is taken to infinity, there have
been several attempts to give an unambiguous renormalization
procedure. The most prominent approach, which was also suggested
by Brown and York, has been to embed the boundary in a reference
space-time, and to subtract the boundary term evaluated in this
space-time. This is similar to the procedure used above to make the
surface charges $J(\xi)$ vanish for anti-de Sitter space. A
problem is, however, that it is not always possible to embed the
boundary in some reference space-time. Therefore, Balasubramanian
and Kraus have suggested an alternative procedure, adding
counterterms to the action that depend only on the boundary metric
to render the energy density finite. The procedure is thus
independent of any reference space-time. 

Let $\gamma_{\mu\nu}$ denote the induced metric of surfaces at
constant $r$. Then, if we add a boundary term and counterterm
action to the Einstein-Hilbert action as
\begin{equation}
S = \frac{1}{16 \pi G} \int_{\mathcal{M}} \sqrt{-g}~(R - 2\Lambda)~d^{d}x - \frac{1}{8\pi G} \int_{\partial \mathcal{M}} \sqrt{-\gamma}~\Theta ~d^{d-1}x + \frac{1}{8 \pi G} S_{ct}(\gamma_{\mu\nu}),
\end{equation}
the stress tensor becomes
\begin{equation}
T_{\mu\nu} = \frac{1}{8\pi G}(\Theta_{\mu\nu} - \Theta \gamma_{\mu\nu} + \frac{2}{\sqrt{-\gamma}}\frac{\delta S_{ct}}{\delta \gamma^{\mu\nu}})
\end{equation}
with $\Theta_{\mu\nu}$ the extrinsic curvature
\begin{equation}
\Theta_{\mu\nu} = \gamma_{\mu\alpha}\gamma_{\nu\beta}D^{\alpha}n^{\beta}
\end{equation}
and $\Theta$ its trace. Now it turns out on the basis of
covariance\ \cite{BK} that for the case of AdS$_3$, the
counterterm should be\footnote{Additional counterterms are
possible, but these will not contribute to the energy-momentum
tensor\ \cite{BK}.}
\begin{equation}
S_{ct} = - \frac{1}{l}\int \sqrt{-\gamma} ~d^{d-1}x,
\end{equation}
so that
\begin{equation}
T_{\mu\nu} = \frac{1}{8\pi G}(\Theta_{\mu\nu}  - \Theta \gamma_{\mu\nu} - \frac{1}{l}\gamma^{\mu\nu}).
\end{equation}
In particular, this causes the stress tensor to vanish for the
Poincar\'e patch of three-dimensional anti-de Sitter space.
Therefore we can look at a deformation of this space in order to
recognize the central charge. Writing the Poincar\'e patch of
AdS$_3$ in terms of $r$ and the coordinates $\tau^+ = t + \phi$, $\tau^- = t -\phi$, we have
\begin{equation}
ds^2 = \frac{l^2}{r^2} dr^2 - r^2 d\tau^{+}d\tau^{-}.
\end{equation}
The asymptotic isometries read
\begin{eqnarray}\label{diffeo}
r &\to& r(1 - \partial_{+} \eta^{+} - \partial_{-}\eta^{-}) \nonumber \\
\tau^{+} &\to& \tau^{+} + 2\eta^{+} + \frac{l^2}{r^2}\partial_{-}^{2}\eta^{-} \nonumber \\
\tau^{-} &\to& \tau^{-} + 2\eta^{-} + \frac{l^2}{r^2}\partial_{+}^{2}\eta^{+},
\end{eqnarray}
and they turn the metric into
\begin{equation}
ds^2 = \frac{l^2}{r^2}dr^2 - r^2d\tau^{+}d\tau^{-} - l^2(\partial_{+}^3 \eta^{+})(d\tau^{+})^2 - l^2(\partial_{-}^3)(d\tau^{-})^2.
\end{equation}
Now we can calculate
\begin{eqnarray}
T_{++} = -\frac{1}{8\pi G l}\gamma_{++} = \frac{l}{8\pi G}\partial_{+}^3 \eta^{+} \nonumber \\
T_{--} = - \frac{1}{8\pi G l}\gamma_{--} = \frac{l}{8 \pi G}\partial_{-}^3 \eta^{-}.
\end{eqnarray}
The transformation rule for the stress tensor under the
diffeomorphisms 
\newline
\mbox{$\tau^{+} \to \tau^{+} + 2\eta^{+}$, $\tau^{-} \to
\tau^{-} + 2\eta^{-}$} is
\begin{eqnarray}
T_{++} \to T_{++} - (4\partial_{+}\eta^{+}T_{++} + 2\eta^{+}\partial_{+}T_{++}) + \frac{c}{12 \pi}\partial_{+}^3 \eta^{+} \nonumber \\
T_{--} \to T_{--} - (4\partial_{-}\eta^{-}T_{--} + 2\eta^{-}\partial_{-}T_{--}) + \frac{c}{12 \pi}\partial_{-}^3 \eta^{-}.
\end{eqnarray}
This is the infinitesimal version of the Schwarzian derivative
that was described in Chapter 2. Because $T_{++}$ and $T_{--}$
were zero before the transformation (\ref{diffeo}), we find
\begin{equation}
c = \frac{3l}{2 G}.
\end{equation}
This central charge corresponds to the one found before in the
Hamiltonian formalism if $T_{++} = \frac{i}{\pi} \sum_{m \in Z}
L_{m} (\tau^{+})^{-m-2}$, as can be checked by means of the
transformation rule given in section 2.3.

\chapter{Conformal Field Theory on the Boundary}

Chapter 6 has revealed a possible analogy between boundary
fluctuations in AdS$_3$ and a conformal field
theory on the $(1+1)$-dimensional boundary. In this chapter we
wish to investigate this analogy further.

Section 7.1 first gives a short introduction to the AdS/CFT correspondence. The original AdS/CFT correspondence concerns an analogy between string theory in a background of AdS$_d$ times some compact manifold, and supersymmetric nonabelian gauge theory in $d-1$ dimensions. In that sense the analogy between fluctuations in AdS$_3$ and a conformal field theory on the boundary is not a full-fledged example of an AdS/CFT correspondence. However, we can use the term in a more obvious sense.

As mentioned before, Coussaert, Henneaux, and van
Driel showed in \cite{Coussaert} that the conformal field
theory living on the boundary of asymptotically AdS$_3$ is Liouville theory. We will give a summary of their derivation. To this end we shall need Chern-Simons theory, which is the subject of section 7.2. The derivation of \cite{Coussaert} is subsequently treated in 7.3.

\section{AdS/CFT Correspondence}

A correspondence between string theory in a $d$-dimensional
anti-de Sitter background (times some compact manifold) and a
gauge theory in $d-1$ dimensions was first noted by Maldacena in
\cite{Maldacena}. The original example concerned the analogy
between the near-horizon limit of $N$ parallel D3-branes in type IIB supergravity, of which the near-horizon geometry is \mbox{AdS$_5 \times$ S$^5$}, and another limit of four-dimensional $\mathcal{N}=4$ super Yang-Mills theory (i.e., nonabelian gauge theory with four independent supersymmetries). This theory has a $U(N)$ gauge symmetry, and the correspondence holds in the limit where both $N$ and $g_{YM}^2 N$ are large, with $g_{YM}$ the Yang-Mills coupling. The symmetries of the supergravity theory are enhanced to the superconformal group near the horizon, similar to the enhancement of the AdS$_3$ symmetries to the conformal group when approaching the boundary. A first test of the correspondence is that the four-dimensional super Yang-Mills theory has the same superconformal symmetry.
Similarly, Maldacena showed that there is a correspondence between branes in type IIB string theory compactified on AdS$_3 \times$ S$^3 \times$ T$^4$
and a $(1+1)$-dimensional conformal field theory. He then conjectured that these analogies continue to hold when moving away from the horizon, which corresponds to moving away from the conformal point on the gauge theory side. The term AdS/CFT correspondence is now more generally used for dualities between
gravity and gauge theories. In this latter sense the correspondence between the Brown-Henneaux boundary fluctuations and $(1+1)$-dimensional Liouville theory may be viewed as an example of an AdS/CFT correspondence.

The AdS/CFT correspondence is intimately tied to the idea of holography, which says that the information contained in a $d$-dimensional theory can sometimes be fully captured in a $(d-1)$-dimensional local field theory. The term holography was coined by G. 't Hooft in 1993 \cite{holography}, and he had several reasons to conjecture such a correspondence.
For one, the fact that the entropy of black holes is proportional to their horizon area suggests that the degrees of freedom of gravity are nonlocal. Another indication that the degrees of freedom of general relativity are not fundamentally $d$-dimensional, is that large volumes with constant energy density collapse into black holes, while they are locally equivalent to small volumes with constant energy density. 

For the standard AdS/CFT correspondence there is another important idea, which is that $U(N)$ gauge theories are equivalent to string theories in the limit of large $N$ (not to be confused with the number of supersymmetries $\mathcal{N}$). This correspondence was also first suggested by 't Hooft \cite{Hooft}. The relevance of this result is that Yang-Mills theory is obtained in the large $N$ limit of a configuration of $N$ parallel D3-branes in type IIB string theory on AdS$_5 \times$S$_5$ that are brought together (the length of the open strings extending between the branes is taken to zero). Maldacena's conjecture is that the full string theory that is obtained when moving away from the black hole horizon corresponds to the theory obtained when the branes are taken back to finite distance.

We will now give a more concrete description of the AdS/CFT correspondence, and see that local quantities in supergravity correspond to super Yang-Mills operators `on the boundary' \cite{Witten again, Gubser}.

A metric for AdS$_5$ (the five-dimensional version of (\ref{warped})) is
\begin{equation}
ds^2 = l^2(dr^2 + e^{2r}(\eta_{\mu\nu}dx^{\mu}dx^{\nu})), 
\end{equation}
where $\eta_{\mu\nu}$ is the four-dimensional Minkowski metric. The
boundary is located at $r = \infty$.
If $\phi_i$ is a free field with mass $m$ propagating in anti-de
Sitter space, with equation of motion
\begin{equation}
(\partial^2 + m^2)\phi = 0,
\end{equation}
there are two linearly independent solutions that are proportional to 
\begin{equation}
e^{-\Delta r},\ \ \ e^{(\Delta - 4)r}
\end{equation}
near the boundary.
Taking the solution that looks near the boundary like
\begin{equation}
\phi_i \sim \phi_i^0 e^{(\Delta - 4)r},
\end{equation}
where
\begin{equation}
\Delta(\Delta -4) = m^2,
\end{equation}
the duality is expressed in the weak-coupling limit of string theory by
\begin{equation}
\exp(-S_{\mathrm{sugra}}(\phi_i)) = \left< \exp\left(\int \phi_i^0
\mathcal{O}_i \right)\right>.
\end{equation}
Here, the supergravity action is evaluated on the classical
solution $\phi_i$, and $\mathcal{O}_i$ represents some operator in
super Yang-Mills theory with conformal dimension $\Delta$. The right-hand side is the generating functional for expectation values in super Yang-Mills theory. 

The AdS/CFT correspondence is a weak/strong coupling duality. On the string theory side there is a dimensionless coupling constant $g_s$, and two parameters with the dimension of length: the curvature radius $l$ of AdS$_5$ and S$^5$, and the string length $l_s$. These are related to the Yang-Mills parameters by
\begin{equation}
g_s = g_{YM}^2,\ \ \ (l/l_s)^4 = 4\pi g_{YM}^2 N \equiv 4\pi\lambda.
\end{equation}
The perturbative expansion for Yang-Mills theory,
\begin{equation}
Z = \sum_{g \ge 0} N^{2 - 2g}f_g(\lambda),
\end{equation}
is valid in the regime where $g_{YM}^2 N$ and $g_{YM}$ are both small, while the expansion for string theory,
\begin{equation}
Z = \sum_{g \ge 0} g_s^{2g-2}Z_g,
\end{equation}
is valid when $g_{YM}$ is small and $g_{YM}^2 N$ is large \cite{de Boer}. 
Testing the correspondence is complicated by the fact that the
strong-coupling regime of neither string theory nor super
Yang-Mills has been completely solved. However, the massless spectrum
provides a good testing ground, and the correspondence has proven to apply at least for massless fields \cite{massless}. The reason that the massless spectrum can be solved is that the massless fields of string theory form shortened, so-called BPS multiplets, of which the conformal dimension is not renormalized. The eigenvalues of the corresponding operators on the gauge theory side are then also not renormalized (they are independent of the coupling). There is of course much more to be said about the AdS/CFT correspondence, but we shall conclude our discussion here and further refer to the literature.

\section{Chern-Simons Theory}

In the following, we shall use the Chern-Simons formulation of $(2+1)$-dimensional gravity with negative cosmological constant. Chern-Simons theory is a topological gauge theory. Therefore, recasting general relativity as a Chern-Simons theory is only possible in three dimensions, where the theory without matter terms has no local propagating degrees of freedom. The role of the gauge group is then played by the space-time isometry group. In general, a Chern-Simons theory exists
in any odd dimension, where it provides a way to describe gauge-invariant mass terms, but not in even dimensions. 

The Chern-Simons Lagrangian in three dimensions for a general compact gauge
group is \cite{Dunne}:
\begin{equation}\label{Chern-Simons}
\mathcal{L}_{CS} = \kappa \epsilon^{\mu\nu\rho}
~\mathrm{Tr}(A_{\mu}\partial_{\nu}A_{\rho} +
\frac{2}{3}A_{\mu}A_{\nu}A_{\rho}) - A_{\mu}J^{\mu},
\end{equation}
where the gauge field $A_{\mu}$ is Lie-algebra valued,
\begin{equation}
A_{\mu} = A_{\mu}^{a}T_{a} 
\end{equation}
with $T_{a}$ the group generators,
and Tr stands for a summation over the internal indices. If the
group is abelian, the fields $A_{\mu}$ commute, and the second term
drops out because of the antisymmetry of the Levi-Civita symbol
$\epsilon^{\mu\nu\rho}$. The equations of motion are
\begin{equation}
\kappa \epsilon^{\mu\nu\rho}F_{\nu\rho} = J^{\mu},
\end{equation}
where the field strength tensor is
\begin{equation}
F_{\mu\nu} =
\partial_{\mu}A_{\nu} - \partial_{\nu}A_{\mu} + [A_{\mu},
A_{\nu}].
\end{equation}
Of course, the last term drops out in the abelian case.
In Lie algebra components:
\begin{equation}
F_{\mu\nu}^{a}T_{a} = \partial_{\mu}A_{\nu}^{a}T_{a} - \partial_{\nu}A_{\mu}^{a}T_{a} + A_{\mu}^{a}A_{\nu}^{b}f_{ab}^{c}T_{c},
\end{equation}
with $f_{ab}^{c}$ the group's structure constants, $[T_{a},T_{b}]
= f_{ab}^{~~c}T_{c}$.
If the current $J^{\mu} = 0$, we obtain the source-free equations
$F_{\mu\nu} = 0$, a demand which is equivalent to the gauge field
$A_{\mu}$ being pure gauge:
\begin{equation}
A_{\mu} = U^{-1}\partial_{\mu}U,
\end{equation}
with $U$ an element of the gauge group.

\section{Liouville Field Theory} 

Chern-Simons theories have been shown to reduce to so-called Wess-Zumino-Witten, or WZW \cite{WZW} theories on the boundary \cite{reduction to WZW}. In particular, the Chern-Simons theory describing $(2+1)$-dimensional gravity with negative cosmological constant reduces under certain boundary conditions to an $SL(2,R)$ WZW model on the cylinder at spatial infinity. In \cite{Coussaert}, it was shown that the Brown-Henneaux conditions are in fact stronger and cause a further reduction to Liouville theory. We will summarize the derivation, and see the nature of the equivalence.

The action for three-dimensional Einstein gravity with negative cosmological constant $\Lambda < 0$ is a sum of two Chern-Simons actions each with gauge group $SL(2,R)$ \cite{Witten}, so that the total gauge group is the product $SL(2,R)_L\times SL(2,R)_R \simeq SO(2,2)$. As before, we should add a term to the action if we are going to take into account surface terms in the variation,
\begin{equation}\label{Chern-Simons action}
S[A,\tilde A]
= S_{CS}[A] -S_{CS}[\tilde A] 
 - \lim_{r \to \infty}\int dt d\phi ~\mathrm{Tr}(A_{\phi}^2) - \lim_{r \to \infty} \int dt d\phi ~\mathrm{Tr}(\tilde A_{\phi}^2),
\end{equation}
where
\begin{equation}
S_{CS}[A] = \int dtdrd\phi ~\mathrm{Tr}(\dot A_r A_{\phi} - \dot A_{\phi}A_r - A_0F_{r\phi}).
\end{equation}
The gauge fields $A_{\mu}$, $\tilde A_{\mu}$ decompose as
\begin{equation}
A_{\mu} =  \omega_{\mu} + \frac{1}{l}e_{\mu},\ \ \ \tilde A_{\mu} = \omega_{\mu} - \frac{1}{l}e_{\mu}.
\end{equation}
Writing out the group index, $e^a_{~\mu}$ is the dreibein field ($g_{\mu\nu} = e^a_{~\mu}e^b_{~\nu}\eta_{ab}$ with $\eta_{ab}$ the flat Minkowski metric), and $\omega^a_{~\mu}$ is the spin connection, which is related to the Riemann tensor by
\begin{eqnarray}
R_{\mu\nu~b}^{~~~a} &=& R_{\mu\nu~\sigma}^{~~~\rho}e^a_{~\rho}e_b^{~\sigma} \nonumber \\
&=& \partial_{\mu}\omega_{\nu~b}^{~a}- \partial_{\nu}\omega_{\mu~b}^{~a}+ [\omega_{\mu}, \omega_{\nu}]^a_{~b}.
\end{eqnarray} 
As in Maxwell theory, the components $A_0$ and $\tilde A_0$ act as Lagrange multipliers, leading to the constraints $F_{r\phi} = \tilde F_{r\phi} = 0$. As mentioned, this amounts to the gauge field being pure gauge. The equations can thus be solved according to\footnote{The derivation is slightly modified for black hole solutions \cite{Coussaert}.}
\begin{equation}\label{pure gauge}
A_i = G_1^{-1}\partial_i G_1,\ \ \ \tilde A_i = G^{-1}_2\partial_i G_2,
\end{equation}
where $i = (r,\phi)$, and $G_1$ and $G_2$ are asymptotically given by
\begin{equation}
G_1 \sim g_1(t,\phi)\mathrm{diag}\left(\sqrt{r}, \frac{1}{\sqrt{r}}\right), \ \ \ G_2 \sim g_2(t,\phi)\mathrm{diag}\left(\frac{1}{\sqrt{r}},\sqrt{r}\right).
\end{equation}
Here, $g_1(t,\phi)$ and $g_2(t,\phi)$ are two arbitrary $SL(2,R)$ group elements.
The particular form (\ref{pure gauge}) for the gauge field causes a reduction of the action (\ref{Chern-Simons action}) to the difference of two chiral WZW actions
\begin{eqnarray}
S^{WZW}_+[g_1] = \lim_{r \to \infty} \int d\phi dt ~\mathrm{Tr}(\dot g_1 \partial_{\phi}g_1 - (\partial_{\phi} g_1)^2) + \int dr d\phi dt ~\mathrm{Tr} (g_1^{-1}dg_1)^3 \nonumber \\
S^{WZW}_-[g_2] = \lim_{r \to \infty} \int d\phi dt ~\mathrm{Tr}(\dot g_2 \partial_{\phi}g_2 - (\partial_{\phi} g_2)^2) + \int dr d\phi dt ~\mathrm{Tr} (g_2^{-1}dg_2)^3
\end{eqnarray}
The combined result turns out to be non-chiral after substituting $g \equiv g_1^{-1}g_2$. 
Besides (\ref{pure gauge}), there are additional conditions following from the Brown-Henneaux asymptotics, which read in terms of $A$ and $\tilde A$, 
\begin{displaymath}
A \sim \left[ \begin{array}{cc}
\frac{dr}{2r} & O\left(\frac{1}{r}\right) \\
rdx^+ & -\frac{dr}{2r}
\end{array} \right]
\end{displaymath}\ \ \ 
\begin{displaymath}
\tilde A \sim \left[ \begin{array}{cc}
 -\frac{dr}{2r} & rdx^- \\
O\left(\frac{1}{r}\right) & \frac{dr}{2r}
\end{array} \right].
\end{displaymath}
Translating these into conditions on $g$ yields
\begin{equation}\label{additional constraints}
(g^{-1}\partial_-g)^{(+)} = 1, \ \ \ (\partial_+g g^{-1})^{(-)} = 1.
\end{equation}
In order to incorporate these extra conditions, the authors of \cite{Coussaert}  Gauss decomposed \\
\begin{equation}
g = \begin{pmatrix} 1 & X \\ 0 & 1 \end{pmatrix} \begin{pmatrix} \exp(\frac{1}{2}\phi) & 0 \\ 0 & \exp(-\frac{1}{2}\phi) \end{pmatrix} \begin{pmatrix} 1 & 0 \\ Y & 1 \end{pmatrix}
\end{equation}
\\
obtaining the action
\begin{equation}\label{WZW action}
S^{WZW} = \int dt d\phi[\frac{1}{2}\partial_+ \phi \partial_-\phi + 2(\partial_-X)(\partial_+Y)\exp(-\phi)].
\end{equation}
There is one last complication arising from the fact the the constraints (\ref{additional constraints}) restrict $\partial_- Y$ and $\partial_+ X$ rather than $X$ and $Y$ themselves. The latter would be appropriate for (\ref{WZW action}), but since the former applies, we need to add another boundary term to the action:
\begin{eqnarray}
S^{WZW}_{impr} = \int dt d\phi[\frac{1}{2}\partial_+ \phi \partial_-\phi + 2(\partial_-X)(\partial_+Y)e^{-\phi}] \nonumber \\
- 2 \oint d\phi(X\partial_+Y + Y \partial_- X)e^{-\phi}|^{t_2}_{t_1}.
\end{eqnarray}
Substituting the additional constraints (\ref{additional constraints}) finally leads to the Liouville action 
\begin{equation}\label{Liouville theory}
S[A, \tilde A] = \int dt d\phi(\frac{1}{2}\partial_+\phi\partial_-\phi + 2e^{\phi}).
\end{equation}
This time, the action does not contain the boundary term that contributed to the central charge of the algebra of Noether charges in the example of section 2.4. Therefore, the analogy has not preserved the central charge of the Virasoro algebra. In fact, the Liouville theory (\ref{Liouville theory}) has an effective central charge $c_{eff} = 1$. In \cite{Myung}, it was proposed that the central charge $c = \frac{3l}{2G}$ is obtained when counting all modes, including those that are non-normalizable, whereas $c = 1$ counts only normalizable modes.
The relation between the Liouville field $\phi$ and the original variables $e^a_{~\mu}$ and $\omega_{~\mu}^a$ is rather complex, but may be traced back through the above derivation. 

It would be interesting to know if analogies such as the one described apply to a wider range of theories. However, a generalization to higher dimensions or theories with matter terms is not straightforward, since in these cases the action can no longer be rewritten as a Chern-Simons action. Therefore, it remains to be seen whether this result can find a wider application.

\chapter{Entropy}

We conclude this thesis with a note on black hole entropy.
There have been successful calculations of black hole entropy in string theory for extreme and near-extreme black holes \cite{Vafa}. Interestingly, Strominger has shown \cite{Strominger} that we can use the \emph{classical} central charge of asymptotically AdS$_3$ to calculate the entropy of the BTZ black hole. The calculation relies on the Cardy formula \cite{Cardy} for the asymptotic density of states, which is a well-known result from conformal field theory. The Cardy formula is specific to two dimensions. It therefore seems that a generalization to other dimensions is only possible if the relevant solution has a special two-dimensional boundary. Although the formula works well for the BTZ black hole, other examples have been found where the correspondence does not hold exactly. A tentative solution to this problem is that Cardy's formula only gives a maximum possible entropy for a given mass \cite{prefactor}. 

There is a well-known analogy between black hole physics and thermodynamics, called black hole thermodynamics. In this analogy, the role of temperature is played by the surface gravity $\kappa$,
\begin{equation}
T_{bh} = \frac{\kappa}{2\pi}.
\end{equation}
Under reasonable circumstances, the horizon of a stationary black hole is a Killing horizon, where Killing vectors become null. If $\chi^{\mu}$ is a normal Killing vector field to the horizon, $\kappa$ is defined by
\begin{equation}
D^{\nu}(\chi^{\mu}\chi_{\mu}) = -2\kappa \chi^{\nu}.
\end{equation}
The entropy, on the other hand, is proportional to the horizon area\footnote{This is how the Bekenstein-Hawking entropy is usually written. Bringing back the appropriate constants, it is actually $S_{bh} =  \frac{A k c^3}{4 h G}$, where $k$ is Boltzmann's constant, $c$ the speed of light, and $h$ Planck's constant.},
\begin{equation}\label{BH}
S_{bh} = \frac{A}{4 G},
\end{equation}
This is the Bekenstein-Hawking entropy \cite{Bekenstein, Hawking}. The analogy suggests that there should be a statistical mechanical interpretation explicitly counting the microstates of the black hole. In $2+1$ dimensions there turns out to be an elegant way to do this, using the central charge of the Virasoro algebra on the boundary. This was first suggested by Strominger in \cite{Strominger}. The method makes use of Cardy's formula for the asymptotic (i.e., large $\Delta$ and $\bar \Delta$) density of states of a conformal field theory with central charge $c$,
\begin{equation}\label{Cardy}
\rho(\Delta, \bar \Delta) \approx \exp\left[2\pi\sqrt{\frac{c\Delta}{6}} + 2\pi\sqrt{\frac{\bar c \bar \Delta}{6}} \right],
\end{equation}
Here, $\Delta$ and $\bar \Delta$ are the eigenvalues of $l_0$ and $\bar l_0$, respectively,
\begin{equation}\label{Delta}
\Delta = \frac{1}{2}(lM + J),\ \ \ \bar \Delta = \frac{1}{2}(lM - J).
\end{equation}
Large values of $\Delta$ and $\bar \Delta$ imply large mass as well as the non-extreme limit $lM \gg J$. 
The entropy can be obtained as the logarithm of the density of states.
Strominger applied this to the BTZ black hole, for which the Bekenstein-Hawking entropy (\ref{BH}) becomes
\begin{equation}
S_{bh} = \frac{\pi\sqrt{16GMl^2 + 8Gl\sqrt{M^2l^2 -J^2}}}{4G}.
\end{equation}
It can be verified that this is the same result as is obtained by substituting (\ref{Delta}) and $c = \bar c = \frac{3l}{2G}$ in (\ref{Cardy}), and the Cardy formula yields the correct density of states for the BTZ black hole. This is a remarkable result, since we have derived the central charge from fully classical considerations. Since the Cardy formula is derived in the context of conformal field theory, the derivation as a whole is semiclassical.

There have been various attempts to apply this method to more general solutions. The approach has been successful for various black hole solutions whose near-horizon geometry is that of the BTZ black hole\footnote{For a list of references, see \cite{maximum entropy}.}. In \cite{maximum entropy}, the three-dimensional Mart\'inez-Zanelli (MZ) black hole, which includes a conformal scalar field and is asymptotically anti-de Sitter, was considered, yielding a result that corresponds to the Bekenstein-Hawking entropy up to a prefactor. The discrepancy is possibly related to the fact that the solution is only one arbitrary member of a class of space-times with the same asymptotics. This makes it debatable to what extent the boundary dynamics belong to this particular solution. It was therefore suggested in \cite{Killing horizons} that the Cardy formula only gives the maximal possible entropy for solutions with mass $M$ and the same asymptotic behavior. The BTZ black hole would then be the solution with maximum entropy, for which the Cardy formula gives the correct result. 

There are many more problems with the approach of Strominger. For one, Cardy's formula is only supposed to work for BTZ black holes with large mass and relatively small angular momentum, but yields the correct entropy for any values of $M$ and $J$.

Moreover, the derivation has assumed a ground state with mass $M = 0$ and angular momentum $J=0$, so that the lowest possible eigenvalues of $l_0$ and $\bar l_0$ become $\Delta_0 = \bar \Delta_0 = 0$. If the lowest Virasoro eigenvalues are different, we instead have to use
\begin{equation}
c_{eff} = c - 24 \Delta_0,\ \ \ \bar c_{eff} = \bar c -24 \bar \Delta_0.
\end{equation}
For instance, using AdS$_3$ ($M=-\frac{1}{8G}$, $J=0$) as a groundstate would yield $c_{eff} = \bar c_{eff} = \frac{6l}{2G}$, and a different value for the entropy.

In addition, it seems mysterious that the Bekenstein-Hawking entropy, which is proportional to horizon area, should be derivable by counting the degrees of freedom on the boundary at infinity. To make the argument more intuitive, Carlip \cite{Killing horizons, near-horizon} presented a near-horizon construction valid for black holes in any dimension, but unfortunately with some seemingly arbitrary boundary conditions. Several possible explanations have been given for the fact that the derivation also works at spatial infinity, although none of them are quite satisfactory. In \cite{Martinec}, a mechanism was suggested by which the central charge may ``flow from the boundary at infinity to the horizon". It has also been proposed that the possibility of a near-infinity calculation is due to the simplicity of three-dimensional gravity. This argument seems to disregard the fact that the first law of black hole thermodynamics relates the entropy to quantities at infinity in any dimension (see appendix). Therefore, the significance of this result remains to be investigated.

Finally, the Bekenstein-Hawking entropy is a formula derived in the context of semiclassical gravity, and may yet be subject to quantum corrections. By improving upon the approximation scheme that led to
(\ref{Cardy}), Carlip \cite{prefactor} calculated a prefactor for Cardy's formula,
\begin{equation}\label{prefactor}
\rho(\Delta, \bar \Delta) \approx  \left(\frac{c}{96\Delta^3} \right)^{1/4}  \left(\frac{\bar c}{96\bar \Delta^3} \right)^{1/4}  \exp\left[2\pi\sqrt{\frac{c\Delta}{6}} + 2\pi\sqrt{\frac{\bar c \bar \Delta}{6}} \right].
\end{equation}
Various near-horizon treatments of diffeomorphisms in the $r-t$ plane in general dimension lead to the following values for $c$ and $\Delta$ \cite{prefactor},
\begin{equation}\label{c}
c = \frac{3A}{2\pi G}\frac{\beta}{\kappa}
\end{equation}
and
\begin{equation}\label{delta}
\Delta = \frac{A}{16\pi G}\frac{\kappa}{\beta},
\end{equation}
where $\beta$ is an undetermined periodicity. Inserting these values
 into the Cardy formula (\ref{Cardy}), this leads to the standard Bekenstein-Hawking entropy. However, inserting it into (\ref{prefactor}) leads to a logarithmic correction to the Bekenstein-Hawking entropy formula,
\begin{equation}
S \sim \frac{A}{4 \hbar G} - \frac{3}{2}\ln\left(\frac{A}{4 \hbar G} \right) + \ldots.
\end{equation}
Taking all these problems into consideration, the subject remains open to many avenues of research.

\chapter{Discussion}

We have reproduced the result of Brown and Henneaux \cite{Brown-Henneaux} that the asymptotic isometry group of AdS$_3$ is extended to the infinite-dimensional conformal group in $1+1$ dimensions when using appropriate boundary conditions. This is a result specific to three dimensions, since the conformal group has only a finite number of generators in dimensions higher than two. The asymptotic isometry group of AdS$_3$ is generated by two copies of the Virasoro algebra with central charge $c = \frac{3l}{2G}$. The central charge shows up at the level of the Poisson brackets and is therefore classical. We have also seen other examples of classical central charges, and shown that there are two types: those that generate a transformation and those that do not. The Brown-Henneaux central charge falls into the latter category. 

The fact that the asymptotic symmetry algebra is the Virasoro algebra with central charge $\frac{3l}{2G}$ leads to the assumption that fluctuations around AdS$_3$ are described by a two-dimensional conformal field theory with the same central charge. We have summarized the derivation of \cite{Coussaert}, in which it was shown that this conformal field theory is Liouville theory. It seems puzzling at first that this theory has an effective central charge $c = 1$. However, a possible resolution has been proposed in \cite{Myung}.

The central charge of asymptotically AdS$_3$ space-times has been used \cite{Strominger} to reproduce the Bekenstein-Hawking entropy of the BTZ black hole through Cardy's formula \cite{Cardy} for the asymptotic density of states. This result is remarkable in two respects. First of all, it suggests that the degrees of freedom of a black hole may not fundamentally live on the black hole horizon. Moreover, it suggests that black hole entropy has a classical origin. Unfortunately, there are some problems with the approach of \cite{Strominger}, which we have discussed. It will be interesting to see whether these issues can be resolved without damaging the original argument. 

\appendix

\chapter[Conserved Quantities]{\huge{Conserved Quantities}}
We give a short characterization of the relation between the surface charges and conserved quantities in general relativity. The surface charges $J(\xi)$ can be written as the integral of an antisymmetric Noether potential $Q^{\mu\nu}(\xi)$ over the boundary at spatial infinity,
\begin{equation}\label{Noether potential}
J(\xi) = \int_{\Sigma^{\infty}_t} \sqrt{h}~ \epsilon_{\mu\nu}Q^{\mu\nu}(\xi) ~d^{d-2}x.
\end{equation}
$Q^{\mu\nu}(\xi)$ derives its name from the fact that, on shell, its derivative is the Noether current belonging to $\xi$,
\begin{equation}
T^{\mu}(\xi) = \partial_{\nu}Q^{\mu\nu}(\xi),
\end{equation}
and $T^{\mu}$ is automatically conserved ($\partial_{\mu}T^{\mu} = 0$) by the antisymmetry of $Q^{\mu\nu}$. However, these conserved currents do not give rise to the usual Noether charges. The conserved quantities are instead obtained from integrals like (\ref{Noether potential}), for only those vector fields $\xi$ that are isometries. Surface charges evaluated on Killing vectors in the time and angular directions give an expression for the total mass and angular momentum of a solution, respectively. If the solution is a black hole, a third conserved quantity, the entropy, can be found by integrating the same expression over a spatial cross section of the Killing horizon,
\begin{equation}\label{expression for entropy}
S = -\int_{\mathrm{\Sigma_{hor}}} \sqrt{h}~\epsilon_{\mu\nu}Q^{\mu\nu}(\xi)~d^{d-2}x \mid_{\xi^{\mu} = 0, \nabla_{[\mu}\xi_{\nu]} = \epsilon_{\mu\nu}}
\end{equation}
where $\epsilon_{\mu\nu}$ is a binormal spanned by two lightlike vectors at the horizon \cite{Bernard} normalized according to $\epsilon^{\mu\nu}\epsilon_{\mu\nu} = -2$. The integral (\ref{expression for entropy}) yields the Bekenstein-Hawking entropy $S_{bh} = \frac{A}{4 G}$ in the case of vacuum general relativity (i.e., $Q^{\mu\nu}$ is the Noether potential belonging to the Einstein-Hilbert action without matter terms).

The first law of black hole thermodynamics,
\begin{equation}
\delta M = \frac{\kappa}{2\pi}\delta S + \Omega \delta J + \Phi dQ + \ldots
\end{equation}
where $\Omega$ is the angular velocity of the horizon, $\Phi$ the electric potential, and $Q$ the electric charge of the black hole, thus relates integrals at infinity with a local quantity at the horizon \cite{Iyer}. This may have a connection with the fact that we were able to derive an expression for the entropy, which seems to live on the horizon, through an analysis conducted at infinity.

\end{document}